\documentclass[aps,pre,twocolumn]{revtex4-1}

\usepackage{palatino}
\usepackage{amsmath}
\usepackage{amssymb}

\usepackage{graphicx}
\usepackage{dcolumn}
\usepackage{boxedminipage}
\usepackage{verbatim}
\usepackage[colorlinks=true,citecolor=blue,linkcolor=blue]{hyperref}

\graphicspath{{figures/}}

\newcommand{\bfm}[1]{{\bf #1}}

\begin{document}

%%%%%%%%%%%%%%%%%%%%%%%%%%%%%%%%%%%%%%%%%%%%%%%%%%%%%%%%%%%%%
% TITLE
%%%%%%%%%%%%%%%%%%%%%%%%%%%%%%%%%%%%%%%%%%%%%%%%%%%%%%%%%%%%%
\title{Replica exchange and expanded ensemble simulations as Gibbs sampling:\\
Simple improvements for enhanced mixing}

\author{John D. Chodera}
\email{jchodera@berkeley.edu}
\affiliation{Research Fellow, California Institute of Quantitative Biosciences (QB3), University of California, Berkeley, 260J Stanley Hall, Berkeley, California 94720, USA}

\author{Michael R. Shirts}
 \thanks{Corresponding author}
 \email{michael.shirts@virginia.edu}
 \affiliation{Department of Chemical Engineering, University of Virginia, VA 22904}

\date{\today}

%%%%%%%%%%%%%%%%%%%%%%%%%%%%%%%%%%%%%%%%%%%%%%%%%%%%%%%%%%%%%
% ABSTRACT
%%%%%%%%%%%%%%%%%%%%%%%%%%%%%%%%%%%%%%%%%%%%%%%%%%%%%%%%%%%%%

\begin{abstract}
The widespread popularity of replica exchange and expanded ensemble algorithms for simulating complex molecular systems in chemistry and biophysics has generated much interest in discovering new ways to enhance the phase space mixing of these protocols in order to improve sampling of uncorrelated configurations. Here, we demonstrate how both of these classes of algorithms can be considered as special cases of Gibbs sampling within a Markov chain Monte Carlo (MCMC) framework.  Gibbs sampling is a well-studied scheme in the field of statistical inference in which different random variables are alternately updated from conditional distributions.
While the update of the conformational degrees of freedom by Metropolis Monte Carlo or molecular dynamics unavoidably generates correlated samples, we show how judicious updating of the thermodynamic state indices---corresponding to thermodynamic parameters such as temperature or alchemical coupling variables---can substantially increase mixing while still sampling from the desired distributions.
We show how state update methods in common use can lead to poor mixing, and present some simple, inexpensive alternatives that can increase mixing of the overall Markov chain, reducing simulation times necessary to obtain estimates of the desired precision.
These improved schemes are demonstrated for several common applications, including an alchemical expanded ensemble simulation, parallel tempering, and multidimensional replica exchange umbrella sampling.

\emph{Keywords: replica exchange simulation, expanded ensemble simulation, the method of expanded ensembles, parallel tempering, simulated scaling, generalized ensemble simulations, extended ensemble, Gibbs sampling, enhanced mixing, convergence rates, Markov chain Monte Carlo (MCMC), alchemical free energy calculations}
\end{abstract}

\maketitle

%%%%%%%%%%%%%%%%%%%%%%%%%%%%%%%%%%%%%%%%%%%%%%%%%%%%%%%%%%%%%
% INTRODUCTION
%%%%%%%%%%%%%%%%%%%%%%%%%%%%%%%%%%%%%%%%%%%%%%%%%%%%%%%%%%%%%

\section{Introduction}
\label{section:introduction}

A broad category of simulation methodologies known as \emph{generalized ensemble}~\cite{okamoto:biopolymers:2001:generalized-ensemble} or \emph{extended ensemble}~\cite{iba:intl-j-mod-phys-c:2001:extended-ensemble} algorithms have enjoyed increasing popularity in the field of biomolecular simulation over the last decade.
The two most popular algorithmic classes within this category are undoubtedly \emph{replica exchange},~\cite{geyer:conference-proceedings:1991:replica-exchange} which includes parallel tempering~\cite{hukushimi-nemoto:j-phys-soc-jpn:1996:parallel-tempering,hansmann:chem-phys-lett:1997:parallel-tempering-monte-carlo,sugita-okamoto:chem-phys-lett:1999:parallel-tempering-md} and Hamiltonian exchange~\cite{sugita-kitao-okamoto:jcp:2000:hamiltonian-exchange,fukunishi-watanabe-takada:jcp:2002:hamiltonian-exchange,jang-shin-pak:prl:2003:hamiltonian-exchange,kwak-hansmann:prl:2005:hamiltonian-exchange}, among others, and its serial equivalent, the method of \emph{expanded ensembles}~\cite{lyubartsev:jcp:1992:expanded-ensembles}, which includes simulated tempering~\cite{marinari-parisi:europhys-lett:1992:simulated-tempering,geyer-thompson:j-am-stat-assoc:1995:expanded-ensembles} and simulated scaling~\cite{li-fajer-yang:jcp:2007:simulated-scaling}.
In both classes of algorithms, a mixture of thermodynamic states are sampled within the same simulation, with each simulation walker able to access multiple thermodynamic states through a stochastic hopping process, which we will generically refer to as consisting of \emph{swaps} or \emph{exchanges}.  
In expanded ensemble simulations, the states are explored via a biased random walk in state space; in replica exchange simulations, multiple coupled walks are carried out in parallel without biasing factors.
Both methods allow estimation of equilibrium expectations at each state as well as free energy differences between states.
In both cases, stochastic transitions between different thermodynamic states can reduce correlation times and increase sampling efficiency relative to straightforward Monte Carlo or molecular dynamics simulations by allowing the system avoid barriers between important configuration substates.

Because of their popularity, these algorithms and their properties have been the subject of intense study over recent years.
For example, given optimal weights, expanded ensemble simulations have been shown to have provably higher exchange acceptance rates than replica exchange simulations using the same set of thermodynamic states~\cite{park:pre:2008:simulated-tempering}.
Higher exchange attempt frequencies have been demonstrated to improve mixing for replica exchange simulations~\cite{sindhikara-meng-roitberg:jcp:2008:exchange-frequency,sindhikara-emerson-roitberg:jctc:2010:exchange-often-and-properly}.
Alternative velocity rescaling schemes have been suggested to improve exchange probabilities~\cite{nadler-hansmann:pre:2007:optimized-replica-exchange-moves}.
Other work has examined the degree to which replica exchange simulations enhance sampling relative to straightforward molecular dynamics simulations~\cite{rhee-pande:biophys-j:2003:multiplexed-replica-exchange,zuckerman-lyman:jctc:2006:replica-exchange-efficiency,gallicchio-levy:pnas:2007:replica-exchange,nymeyer:jctc:2008:replica-exchange-efficiency,tavan:cpl:2008:pseudoconvergence,rosta-hummer:jcp:2009:replica-exchange-efficiency,rosta-hummer:jcp:2010:simulated-tempering-efficiency}.
Numerous studies have examined the issue of how to optimally choose thermodynamic states to enhance sampling in systems with second-order phase transitions~\cite{kofke:2002:jcp:acceptance-probability,katzberger-trebst-huse-troyer:j-stat-mech:2006:feedback-optimized-parallel-tempering,trebst-troyer-hansmann:jcp:2006:optimized-replica-selection,nadler-hansmann:pre:2007:generalized-ensemble,gront-kolinski:j-phys-condens-matter:2007:optimized-replica-selection,park-pande:pre:2007:choosing-weights-simulated-tempering,shenfeld-xu:pre:2009:thermodynamic-length}, though systems with strong first-order-like phase transitions (such as two-state protein systems) remain challenging~\cite{neuhaus-magiera-hansmann:pre:2007:parallel-tempering-first-order,straub:jcp:2010:generalized-replica-exchange}.
A number of combinations~\cite{fenwick-escobedo:jcp:2003:replica-exchange-expanded-ensembles,mitsutake-okamoto:jcp:2004:rest} and elaborations~\cite{mitsutake-sugita-okamoto:2003:remuca,rhee-pande:biophys-j:2003:multiplexed-replica-exchange,simmerling:jctc:2007:reservoir-replica-exchange,gallicchio-levy-parashar:j-comput-chem:2008:asynchronous-replica-exchange,hansmann:physica-a:2010:replica-exchange} of these algorithms have also been explored. 
A small number of publications have examined the mixing and convergence properties of replica exchange and expanded ensemble algorithms with mathematical rigor~\cite{madras-randall:annals-appl-prob:2002:decomposition-theorem,bhatnagar-randall:acm:2004:torpid-mixing,woodard_conditions_2009,woodard_sufficient_2009}, but there remain many unanswered questions about these sampling algorithms at a deep theoretical level.

Standard practice for expanded ensemble and replica exchange simulations is that exchanges are to be attempted only between ``neighboring'' thermodynamic states---for example, the states with temperatures immediately above or below the current temperature in a simulated or parallel tempering simulation~\cite{hukushimi-nemoto:j-phys-soc-jpn:1996:parallel-tempering,hansmann:chem-phys-lett:1997:parallel-tempering-monte-carlo,sugita-okamoto:chem-phys-lett:1999:parallel-tempering-md,sugita-kitao-okamoto:jcp:2000:hamiltonian-exchange,fukunishi-watanabe-takada:jcp:2002:hamiltonian-exchange,jang-shin-pak:prl:2003:hamiltonian-exchange,kwak-hansmann:prl:2005:hamiltonian-exchange}. 
The rationale behind this choice is that states further away in state space will have low probability of acceptance due to diminished phase space overlap, and thus attempts should focus on the states for which exchange attempts are most likely to be accepted.
Increasing the proximity of neighboring thermodynamic states in both kinds of simulations can further increase the probability that exchange attempts will be accepted. However, restricting exchange attempts to neighboring states can then result in slow overall diffusion in state space due to the larger number of replicas needed to span the thermodynamic range of interest~\cite{machta:pre:2009:parallel-tempering}.
Some exchange schemes have been proposed to improve this diffusion process, such as all-pairs exchange~\cite{izaguirre:jcp:2007:all-pairs-exchange}, and optimized exchange moves~\cite{nadler-hansmann:pre:2007:optimized-replica-exchange-moves} but the problem is still very much a challenge (see Ref.~\cite{tavan:cpl:2009:exchange-schemes} for a recent comparison).
The problem of slow diffusion is exacerbated in ``multidimensional'' simulations that make use of a 2D or 3D grid of thermodynamic states~\cite{sugita-kitao-okamoto:jcp:2000:hamiltonian-exchange,paschek-garcia:prl:2004:temperature-pressure-replica-exchange,jiang_free_2010}, where diffusion times in state space increase greatly due to the increase in dimensionality~\cite{rudnick_elements_2010}.

Here, we show how the many varieties of expanded ensemble and replica exchange simulations can all be considered to be forms of \emph{Gibbs sampling}, a sampling scheme well-known to the statistical inference literature
~\cite{geman-geman:1984:gibbs-sampling,jun-s-liu:mcmc}, though unrelated to simulations in the ``Gibbs ensemble'' for determining phase equilibria~\cite{panagiotopoulos:mol-phys:1987:gibbs-ensemble,panagiotopoulos:mol-phys:1988:gibbs-ensemble,footnote1}. 
When viewed in this statistical context, a number of alternative schemes can readily be proposed for updating the thermodynamic state while preserving the distribution of configurations and thermodynamic states sampled by the algorithm.  
By making simple modifications to the exchange attempt schemes, we show that great gains in sampling efficiency can be achieved under certain conditions with little or no extra cost.  
There is essentially no drawback to implementing these algorithmic improvements, as the additional computational cost is negligible, their implementation sufficiently simple to encourage widespread adoption, and there appears to be no hindrance of sampling in cases where these schemes offer no great efficiency gain.
Importantly, we also demonstrate that schemes that encourage mixing in state space can also encourage mixing of the overall Markov chain, reducing correlation times in coordinate space, leading to more uncorrelated samples being generated for a fixed amount of computer time.  

This paper is organized as follows.
In \emph{Theory} (Section~\ref{section:theory}), we describe expanded ensemble and replica exchange algorithms in a general way, casting them as a form of Gibbs sampling.
{In \emph{Algorithms} (Section~\ref{section:algorithms}), we propose multiple approaches to the state exchange process in both classes of algorithm with the aim of encouraging faster mixing in among the thermodynamic states accessible in the simulation, and hence the overall Markov chain.}
In \emph{Illustration} (Section~\ref{section:illustration}), we illustrate how and why these modified schemes enhance mixing of the overall chain for a simple one-dimensional model system.
In \emph{Applications} (Section~\ref{section:applications}), we apply these algorithmic variants to some examples from physical chemistry, using several different common benchmark systems from biomolecular simulation, and examine several metrics of simulation efficiency.
Finally, we make recommendations for the adoption of simple algorithmic variants that will improve efficiency in \emph{Discussion} (Section~\ref{section:discussion}).

%%%%%%%%%%%%%%%%%%%%%%%%%%%%%%%%%%%%%%%%%%%%%%%%%%%%%%%%%%%%%
% THEORY
%%%%%%%%%%%%%%%%%%%%%%%%%%%%%%%%%%%%%%%%%%%%%%%%%%%%%%%%%%%%%

\section{Theory}
\label{section:theory}

Before describing our suggested algorithmic modifications (\emph{Algorithms}, Section~\ref{section:algorithms}), we first present some theoretical tools we will use to analyze expanded ensemble and replica exchange simulations in the context of Gibbs sampling.

\subsection{Thermodynamic states and thermodynamic ensembles}

To be as general as possible, we describe the expanded ensemble and replica exchange algorithms as sampling a mixture of $K$ thermodynamic states.
Here, a \emph{thermodynamic state} is parameterized by a vector of time-independent thermodynamic parameters $\lambda$.
For notational convenience and to make what follows general, we define the \emph{reduced potential}~\cite{shirts-chodera:jcp:2008:mbar} $u(x)$ of a physical system,
\begin{eqnarray}
u(x) &=& \beta \left[ H(x) + p V(x) + \sum_i\mu_i n_i(x) + \cdots \right], \label{equation:reduced-potential}
\end{eqnarray}
corresponding to its thermodynamic state, 
where $x$ denotes the configuration of the system specifying any physical variables allowed to change, including the volume $V(x)$ (in the case of a constant pressure ensemble) and $n_i(x)$ the number of molecules of each of $i=1,\ldots,M$ components of the system, in the case of a (semi)grand ensemble.  
The reduced potential is a function of the Hamiltonian $H$, the inverse temperature $\beta = (k_B T)^{-1}$, the pressure $p$, and the vector of chemical potentials for each of $M$ components $\mu_i$. 
Other thermodynamic parameters and their conjugate coordinates can be included in a similar manner, or some of these can be omitted, as required by the physics of the system.  We denote the set of all thermodynamic parameters by $\lambda  \equiv \{\beta, H, p, \vec{\mu}, \ldots\}$. 

We next denote a configuration of the molecular system by $x \in \Omega$, where $\Omega$ is allowed configuration space, which may be continuous or discrete.   A choice of thermodynamic state gives rise to set of configurations of the system that are sampled by a given time-independent probability distribution at equilibrium. 
So each $x$ will have associated unnormalized probability density $q(x)$, which is a function of $\lambda$, where $q(x) > 0$ for all $x \in \Omega$. If we define the normalization constant, or \emph{partition function}, $Z$ as:
\begin{eqnarray}
Z &\equiv& \int_\Omega dx \, q(x) 
\end{eqnarray}
we can define a normalized probability density
\begin{eqnarray}
\pi(x) &=& Z^{-1} \, q(x).
\end{eqnarray}

A physical system in equilibrium with its environment obeying classical statistical mechanics will sample configurations distributed according to the Boltzmann distribution,
\begin{eqnarray}
q(x) &\equiv& e^{-u(x)} .
\end{eqnarray}
In this paper, we consider a set of $K$ thermodynamic states defined by their thermodynamic parameter vectors, $\lambda_k  \equiv \{\beta_k, H_k, p_k, \vec{\mu}_k, \ldots\}$, with $k = 1,\ldots,K$, where $H_k$ denotes any modifications of the Hamiltonian $H$ as a function of $k$, including biasing potentials.  Each new choice of $k$ gives rise to a reduced potential $u_k$, unnormalized and normalized probability distributions $q_k(x)$ and $\pi(x,k)$, and a partition function $Z_k$. 
Although in this paper, we generally assume a Boltzmann distribution, there is nothing to prevent some or all of the states from being sampled using non-thermodynamic (non-Boltzmann) statistics using alternative choices of the unnormalized density $q_k(x)$, as in the case of multicanonical simulations~\cite{mezei:j-comp-phys:1987:muca} or Tsallis statistics~\cite{tsallis:j-stat-phys:1988:tsallis-statistics}.  To ensure that any configuration $x$ has finite, nonzero density in all $K$ thermodynamic states, we additionally require that the same thermodynamic parameters be specified for all thermodynamic states, though their values may of course differ.

\subsection{Gibbs sampling}

Suppose we wish to sample from the joint distribution of two random variables, $x$ and $y$. 
We denote this joint distribution by $\pi(x,y)$.
Often, it is not possible to directly generate uncorrelated sample pairs $(x,y)$ from the joint distribution due to the complexity of the function $\pi(x,y)$.
In these cases, a standard approach to sampling is to use some form of Markov chain Monte Carlo (MCMC)~\cite{jun-s-liu:mcmc}, such as the Metropolis-Hastings algorithm~\cite{metropolis:jcp:1953:metropolis-monte-carlo,hastings:biometrika:1970:metropolis-hastings} or hybrid Monte Carlo~\cite{duane:1987:phys-lett-b:hybrid-monte-carlo}.
While general in their applicability, MCMC algorithms suffer from the drawback that they often must generate \emph{correlated} samples,  potentially requiring long running times to produce a sufficient number of effectively uncorrelated samples to allow the computation of properties to the desired precision~\cite{mueller-krumbhaar:j-stat-phys:1973:monte-carlo,janke:2002:statistical-error}.

Assume we can draw samples, either independently or through some Markov chain Monte Carlo procedure, from the \emph{conditional} distributions of one or more of the variables, $\pi(x | y)$ or $\pi(y | x)$, where the value of the second variable is fixed.
To generate a set of sample pairs $\{(x^{(1)}, y^{(1)}), \, (x^{(2)}, y^{(2)}), \ldots\}$ from $\pi(x,y)$, we can iterate the update scheme:
\begin{eqnarray}
x^{(n+1)} | y^{(n)} \:\:\:\:\: &\sim& \pi(x | y^{(n)}) \nonumber \\
y^{(n+1)} | x^{(n+1)} &\sim& \pi(y | x^{(n+1)}) \nonumber
\end{eqnarray}
where $x \sim \pi$ denotes that the random variable $x$ is sampled or ``updated'') from the distribution $\pi(x)$.

This procedure is termed \emph{Gibbs sampling} or \emph{the Gibbs sampler} in the statistical literature, and has been employed and studied extensively~\cite{geman-geman:1984:gibbs-sampling,jun-s-liu:mcmc}.
In many cases, it may be possible to draw uncorrelated samples from either or both distributions, but this is not required~\cite{footnote2}.
The choice of which variable to update---in this example, $x$ or $y$---can be either deterministic (e.g.~update $x$ then $y$) or stochastic (e.g.~a random number determines whether $x$ or $y$ is to be updated); both schemes sample from the desired joint distribution $\pi(x,y)$. However, each method has different dynamic properties and can introduce different correlation structure in the sequence of sample pairs. In particular, we note that a stochastic choice of which variable to update obeys detailed balance, while a deterministic choice obeys the weaker balance condition~\cite{deem:jcp:1999:balance}.  
In both cases, the distribution $\pi(x,y)$ is preserved.   

In the sections below, we describe how expanded ensemble and replica exchange simulations can be considered as special cases of Gibbs sampling on the probability distribution $\pi(x,k)$, which is now a function of both coordinates and thermodynamic states, and how this recognition allows us to consider simple variations of these techniques that will enhance mixing in phase space with little or no extra cost.  
In the algorithms we consider here, the thermodynamic state variable $k$ is discrete, but continuous $k$ are also completely valid in this formalism.
v
\subsection{Expanded ensembles}

In an expanded ensemble simulation~\cite{lyubartsev:jcp:1992:expanded-ensembles}, a single replica or ``walker'') samples pairs $(x,k)$ from a joint distribution of configurations $x \in \Gamma$ and state indices $k \in \{1,\ldots,K\}$ given by,
\begin{eqnarray}
\pi(x,k) &\propto& \exp[-u_k(x) + g_k] ,
\end{eqnarray}
where $g_k$ is an state-dependent weighting factor. This space is therefore a {\em mixed}, {\em generalized}, or {\em expanded} ensemble which samples from multiple thermodynamic ensembles simultaneously. $g_k$ is chosen to give a specific weighting of each subensemble in the expanded ensemble, and is generally determined through some iterative procedure~\cite{lyubartsev:jcp:1992:expanded-ensembles,marinari-parisi:europhys-lett:1992:simulated-tempering,wang-landau:prl:2001:wang-landau,park-ensign-pande:pre:2006:bayesian-weight-update,park-pande:pre:2007:choosing-weights-simulated-tempering,li-fajer-yang:jcp:2007:simulated-scaling,chelli:jctc:2010:optimal-weights-expanded-ensembles}. The set of $g_k$ is frequently chosen to give each thermodynamic ensemble equal probability, in which case $g_k=-\ln Z_k$, but they can be set to arbitrary values as desired.

In the context of Gibbs sampling, an expanded ensemble simulation proceeds by alternating between sampling from the two conditional distributions,
\begin{eqnarray}
\pi(x | k) &=& \frac{q_k(x)}{\int_\Omega dx \, q_k(x)} = \frac{e^{-u_k(x)}}{\int_\Omega dx \, e^{-u_k(x)}} \\
\pi(k | x) &=& \frac{e^{g_k}q_k(x)}{\sum\limits_{k'=1}^K e^{g_{k'}}q_{k'}(x)} = \frac{e^{g_k - u_k(x)}}{\sum\limits_{k'=1}^K e^{g_{k'} - u_{k'}(x)}} .\label{equation:expanded-ensemble-gibbs-update}
\end{eqnarray}
In all but trivial cases, sampling from the conditional distribution $\pi(x | k)$ must be done using some form of Markov chain Monte Carlo sampler that generates correlated samples, due to the complex form of $u_k(x)$ and the difficulty of computing the normalizing constant in the denominator~\cite{jun-s-liu:mcmc}. Typically, Metropolis-Hastings Monte Carlo~\cite{metropolis:jcp:1953:metropolis-monte-carlo,hastings:biometrika:1970:metropolis-hastings} or molecular dynamics is used~\cite{footnote3}, generating an updated configuration $x^{(n+1)}$ that is correlated with the previous configuration $x^{(n)}$.
However, as we will see in Algorithms (Section~\ref{section:algorithms-expanded-ensemble}), multiple choices for sampling from the conditional distribution $\pi(k | x)$ are possible due to the simplicity of its form.

\subsection{Replica exchange ensembles}
\label{section:replica-exchange-ensembles}
In a replica exchange, we consider $K$ simulations, with one simulation in each of the thermodynamic $K$ states.  The current state of the replica exchange simulation is given by $(X,S)$, where $X$ is a vector of the replica configurations, $X \equiv \{x_1, x_2, \ldots, x_K\}$, and $S \equiv\{s_1,\ldots,s_K\} \in \mathcal{S}_K$ is a permutation of the state indices $S \equiv \{1, \ldots, K\}$ associated with each of the replica configurations $X \equiv \{x_1, \ldots, x_K\}$. Then:
\begin{eqnarray}
\pi(X, S) &\propto& \prod_{i=1}^{K} q_{s_i}(x_i) \propto \exp\left[-\sum_{i=1}^K u_{s_i}(x_i)\right]
\label{eq:parallereplica}
\end{eqnarray}
with the conditional densities therefore given by
\begin{eqnarray}
\pi(X | S) &=& \prod_{i=1}^K \left[ \frac{e^{-u_{s_i}(x_i)}}{\int_\Omega dx \, e^{-u_{s_i}(x_i)}}\right] \\
\pi(S | X) &=& \frac{\exp\left[- \sum\limits_{i=1}^K u_{s_i}(x_i) \right]}{\sum\limits_{S' \in \mathcal{S}_K} \exp\left[- \sum\limits_{i=1}^K u_{s'_i}(x_i) \right]}
\end{eqnarray}
As in the case of expanded ensemble simulations, updating of configurations $X$ must be by some form Markov chain Monte Carlo or molecular dynamics simulation, invariably generating configurations with some degree of correlation. Unlike the case of expanded ensembles, generating independent samples in the conditional permutation space is very challenging for even moderate numbers of states because of the expense of computing the denominator of $\pi(S | X)$~\cite{footnote4}, which includes a sum over all permutations in the set $\mathcal{S}_K$. 
However, as we shall see in Section~\ref{section:algorithms-replica-exchange}, there are still effective ways to generate \emph{nearly} uncorrelated permutations that have improved mixing properties over traditional exchange attempt schemes.

%%%%%%%%%%%%%%%%%%%%%%%%%%%%%%%%%%%%%%%%%%%%%%%%%%%%%%%%%%%%%
% ALGORITHMS
%%%%%%%%%%%%%%%%%%%%%%%%%%%%%%%%%%%%%%%%%%%%%%%%%%%%%%%%%%%%%

\section{Algorithms}
\label{section:algorithms}

We now describe the \emph{algorithms} used in sampling from the expanded ensemble and replica exchange ensembles described in \emph{Theory} (Section~\ref{section:theory}).
We start with the typical neighbor exchange schemes commonly used in the literature, and then describe additional novel schemes based on Gibbs sampling that can encourage more rapid mixing among the accessible thermodynamic states.

\subsection{Expanded ensemble simulation}
\label{section:algorithms-expanded-ensemble}

For an expanded ensemble simulation, the \emph{conditional} distribution of the state index $k$ given $x$ is, again:
\begin{eqnarray}
\pi(k | x) &=& \frac{e^{g_k - u_k(x)}}{\sum\limits_{k'=1}^K e^{g_{k'} - u_{k'}(x)}} . \nonumber
\end{eqnarray}
We can use any proposal/acceptance scheme that ensures this conditional distribution is sampled in the long run for any fixed $x$. 
We can choose at each step to sample in $k$ or $x$ depending according to some fixed probability $p$, in which case detailed balance is obeyed.  We can also alternate $N_k$ and $N_x$ steps of $k$ and $x$ sampling, respectively.  Although this algorithm does not satisfy detailed balance, it does satisfy the weaker condition of \emph{balance}~\cite{deem:jcp:1999:balance} which is sufficient to preserve sampling from the joint stationary distribution $\pi(x,k)$.  In the cases that proposal probabilities are based on past history however, the algorithm will not preserve the equilibrium distribution~\cite{reinhardt:cpl:2000:step-size-adjustment}).

\subsubsection{Neighbor exchange}
\label{section:algorithms:expanded-ensemble:neighbor-exchange}

In the neighbor exchange scheme, the proposed state index $j$ given the current state index $i$ is chosen randomly from one of the neighboring states, $i \pm 1$, with probability,
\begin{eqnarray}
\alpha(j | x, i) &=& \begin{cases}
\frac{1}{2} & \mathrm{if}\:\:j = i-1 \\
\frac{1}{2} & \mathrm{if}\:\:j = i+1 \\
0 & \mathrm{else}
\end{cases}
\label{eq:replica-up-and-down}
\end{eqnarray}
and accepted with probability,
\begin{eqnarray}
\lefteqn{P_\mathrm{accept}(j | x, i) =} \nonumber \\
&&\hspace{0.3in}\mbox{} \begin{cases}
0 & \mathrm{if}\:\:j \notin \{1, \ldots, K\} \\
\min\left\{1, \frac{e^{g_{j} - u_{j}(x)}}{e^{g_{i} - u_{i}(x)}} \right\} & \mathrm{else} \\
\end{cases}
\label{eq:mc-with-ends}
\end{eqnarray}
This scheme was originally suggested by Marinari and Parisi~\cite{marinari-parisi:europhys-lett:1992:simulated-tempering} and has been used extensively in subsequent work~\cite{hansmann-okamoto:1996:pre:simulated-tempering,fenwick-escobedo:jcp:2003:replica-exchange-expanded-ensembles}. 
A slight variation of this scheme considers the set $\{1, \ldots, K\}$ to lie on a torus, such that state $i + n K$ is equivalent to state $i$ for integral $n$, with the proposal and acceptance probability otherwise left unchanged.

An alternative scheme avoids having to reject choices of $j$ that lead to $j \notin \{1, \ldots, K\}$ by modifying the proposal scheme,
\begin{eqnarray}
\alpha(j | x, i) &=& \begin{cases}
\frac{1}{2} & \mathrm{if}\:\:k \in \{2,\ldots,K-1\}, |j-i| = 1 \\
1 & \mathrm{if}\:\:i=1,j=i+1 \le K\\
1 & \mathrm{if}\:\:i=K,j=i-1 \ge 1\\
0 & \mathrm{else}
\end{cases}
\end{eqnarray}
and modifying the acceptance criteria for these two moves to be~\cite{fenwick-escobedo:jcp:2003:replica-exchange-expanded-ensembles}
\begin{eqnarray}
P_\mathrm{accept}(j | x, i) &=& \min\left\{1, \frac{1}{2} \frac{e^{g_{j} - u_{j}(x)}}{e^{g_{i} - u_{i}(x)}} \right\}\end{eqnarray}
 to include the correct Metropolis-Hastings ratio of proposal probabilities.

\subsubsection{Independence sampling}
\label{section:algorithms:expanded-ensemble:gibbs-sampling}

The most straightforward way of generating an uncorrelated state index
$i$ from the conditional distribution $\pi(k | i)$ is by
\emph{independence sampling}, in which we propose an update of the state index $i$
by drawing a new $j$ from $\pi(i | x)$ with probability
\begin{eqnarray}
\alpha(j | x, i) &=& \pi(i | x) 
\end{eqnarray}
and always accepting this new $j$. 
While well-known in the statistical inference literature~\cite{jun-s-liu:mcmc}---and the update scheme most closely associated with the use of the Gibbs sampler there---this scheme has been recently discovered independently in the context of molecular simulation~\cite{rosta-hummer:jcp:2010:simulated-tempering-efficiency}. 
A straightforward way to implement this update scheme is to generate a uniform random number $r$ on the interval $[0,1)$, and select the smallest $k$ where $r < \sum_{i=1}^k \pi(i|x)$.

\subsubsection{Metropolized independence sampling}
\label{section:algorithms:expanded-ensemble:metropolized-gibbs}

In what we term a \emph{Metropolized independence sampling} move~\cite{liu:biometrika:1996:metropolized-gibbs}, a new state index $k'$ is proposed from the distribution,
\begin{eqnarray}
\alpha(j | x, i) &=& \begin{cases}
\frac{\pi(j | x, i)}{1 - \pi(j | x, i)} & j \ne i \\
0 & j = i
\end{cases}
\end{eqnarray}
and accepted with probability,
\begin{eqnarray}
P_\mathrm{accept}(j | x, i) &=& \min\left\{ 1, \frac{1 - \pi(i | x,i)}{1 - \pi(j | x,i)} \right\} .
\end{eqnarray}
This scheme has the surprisingly property that, despite including a rejection step (unlike the independence sampling in Section~\ref{section:algorithms:expanded-ensemble:gibbs-sampling} above), the mixing rate in $\pi(k | x)$ can be proven to be greater than that of independence sampling~\cite{liu:biometrika:1996:metropolized-gibbs}, using the same arguments that Peskun used to demonstrate the optimality of the Metropolis-Hastings criteria over other criteria for swaps between two states.  
This can be rationalized by noting that Metropolized independence sampling updates will always try move away from the current state, whereas standard independence sampling has some nonzero probability to propose to remain in the current state.

\subsubsection{Restricted range sampling}
\label{section:algorithms:expanded-ensemble:restricted-range-gibbs}

In some situations, such as simulated scaling~\cite{li-fajer-yang:jcp:2007:simulated-scaling} or other schemes in which the Hamiltonian differs in a non-trivial way among thermodynamic states, there may be a non-negligible cost in evaluating the unnormalized probability distributions $q_k(x)$ for all $k$. 
Because transitions to a states with minimal phase space overlap will have very low probability, prior knowledge of which states have the highest phase space overlap could reduce computational effort with little loss in sampling efficiency if states with poor overlap are excluded from consideration for exchange. 

One straightforward way to implement such a \emph{restricted range sampling} scheme is to define a set of proposal states $\mathcal{S}_i$ for each state $i \in \{1, \ldots, K\}$, with the requirement that $i \in \mathcal{S}_j$ if and only if $j \in \mathcal{S}_i$, and propose transitions from the current $(x,i)$ to a new state $j$ with probability,
\begin{eqnarray}
\alpha(j | x, i) &=& \begin{cases}
\frac{e^{g_j - u_j(x)}}{\sum\limits_{k \in S_i} e^{g_k - u_k(x)}} & j \in \mathcal{S}_i \\
\hspace{0.8cm} 0 & j \notin \mathcal{S}_i
\end{cases} .
\end{eqnarray}
This proposal is accepted with probability,
\begin{eqnarray}
P_{\text{accept}}(j | x, i) &=& \min\left(1,\frac{\sum\limits_{k \in S_i} e^{g_k - u_k(x)}}{\sum\limits_{k' \in S_j} e^{g_{k'} - g_{k'}(x)}}\right). \label{equation:restricted-range-acceptance-criteria}
\end{eqnarray}

We can easily see that this scheme satisfies detailed balance for fixed $x$.
The probability the sampler is initially in $i \in \mathcal{S}_j$ and transitions to $j \in \mathcal{S}_i$, where $j \ne i$, is given by,
\begin{eqnarray}
\lefteqn{\pi(i | x)\alpha(j | x, i)P_{\text{accept}}(j | x, i)}  \nonumber \\
&=& \left[\frac{e^{g_i - u_i(x)}}{Z(\mathcal{S}_{\text{all}})}\right]\left[\frac{e^{g_j - u_j(x)}}{Z(\mathcal{S}_i)}\right] \left[\min\left(1,\frac{Z(\mathcal{S}_i)}{Z(\mathcal{S}_j)}\right)\right] \\
&=& \left[\frac{e^{g_j - u_j(x)}e^{g_i - u_i(x)}}{Z(\mathcal{S}_{\text{all}})}\right]\left[\min\left(Z^{-1}(\mathcal{S}_i),Z^{-1}(\mathcal{S}_j)\right)\right]\\
&=& \left[\frac{e^{g_j - u_j(x)}}{Z(\mathcal{S}_{\text{all}})}\right]\left[\frac{e^{g_i - u_i(x)}}{Z(\mathcal{S}_j)}\right] \left[\min\left(1,\frac{Z(\mathcal{S}_j)}{Z(\mathcal{S}_i)}\right)\right] \\
&=& \pi(j | x)\alpha(i | x, j)P_{\text{accept}}(i | x, j) 
\end{eqnarray}
where $Z(\mathcal{S}_i) = \sum_{k \in S_i} e^{g_k - u_k(x)}$, and $\mathcal{S}_{\text{all}} = \{1,\ldots,K\}$.
This is simply the detailed balance condition, ensuring that this scheme will sample from the distribution $\pi(i | x)$.
Therefore, this scheme samples from the stationary probability $\pi(j | x)$.

For example, we can define $\mathcal{S}_i = \{i-n,\ldots,i+n\}$, with $n \ll K$, for all $i$, making appropriate adjustments to this range at $i < n$ and $i > K-n$.  
Then we only need to compute the reduced potentials for states $\{ \min(1, i-2n),\ldots, \max(K,i+2n) \}$, rather than all states $\{1, \ldots, K\}$.  
The additional evaluations for $\{ \min(1,i-2n),\ldots,i-n-1 \}$ and $\{ \max(i+n+1,K) \ldots,\max(K,i+2n) \}$ are required to ensure that we can calculate both sums in the acceptance criteria (Eq.~\ref{equation:restricted-range-acceptance-criteria}).

Restricted range sampling simply reduces to independence sampling, as presented in Section~\ref{section:algorithms:expanded-ensemble:gibbs-sampling}, when $\mathcal{S}_i = \{1,\ldots,K\}$, and all proposals are therefore accepted. We also note that Metropolized independence sampling, in Section~\ref{section:algorithms:expanded-ensemble:metropolized-gibbs} is exactly equivalent to using the restricted range scheme with $\mathcal{S}_i = \{1,\ldots,K\}$ \emph{excluding} $i$, such that $\alpha(i|x,i) =0$ for all $i$.  
%MRS: setminus might confuse some people
Any other valid scheme of sets $\mathcal{S}_i$ can be Metropolized by removing $i$ from $\mathcal{S}_i$.

%Finally, standard nearest neighbor Metropolis exchange can be seen as a restricted range scheme where the sets $\mathcal{S}_i$ are not fixed, but are instead stochastically determined at each step randomly.  With equal probability, we have $\mathcal{S}_i = \{i,i+1\}$ or $\mathcal{S}_i = \{i-1,i\}$.  
%MRS: Is this worth including after all?

Clearly, other state decomposition schemes exist, though the efficiency of such schemes will almost certainly depend on the underlying nature of the thermodynamic states under study.  It is possible to define state schemes that preserve detailed balance, but that are not ergodic, such as $\mathcal{S}_1=\mathcal{S}_3=\mathcal{S}_5=\{1,3,5\}$ and $\mathcal{S}_2=\mathcal{S}_4=\mathcal{S}_6=\{2,4,6\}$ for $K=6$, so some care must be taken.  In most cases, users will likely use straightforward rules to find locally defined sets such as $\mathcal{S}_i = \{i-n,\ldots,i+n\}$ or the Metropolized version $\mathcal{S}_i = \{i-n,\ldots,i-1,i+1,\ldots,i+n\}$, and ergodicity as well as detailed balance will be satisfied.  Further analysis of the performance tradeoffs involved in choices of the sets, situations where sets might be chosen stochastically, or more efficient choices of sets that satisfy only balance is beyond the scope of this study.

%Since restricting the range of sampling lessens the possibility of visiting new states, it is likely that most of these further restricted range variations will likely be less efficient unless the cost of computing all of the $q_k(x)$ is large.

\subsubsection{Other schemes}

The list above is by no means intended to be exhaustive---many other schemes can be used for updating the state index $k$, provided they sample from $\pi(k | x)$.
Compositions of different schemes are also permitted---even something simple as application of the neighbor exchange scheme a number of times, rather than just once, could potentially improve mixing properties at little or no additional computational cost.

%%%%%%%%%%%%%%%%%%%%%%%%%%%%%%%%%%%%%%%%%%%%%%%%%%%%%%%%%%%%%
\subsection{Replica exchange simulation}
\label{section:algorithms-replica-exchange}

\subsubsection{Neighbor exchange}
\label{section:algorithms:replica-exchange:neighbor-exchange}

In standard replica exchange simulation algorithms, an update of the state permutation $S$ of the $(X,S)$ sampler state only considers exchanges between neighboring states~\cite{hukushimi-nemoto:j-phys-soc-jpn:1996:parallel-tempering,hansmann:chem-phys-lett:1997:parallel-tempering-monte-carlo,sugita-okamoto:chem-phys-lett:1999:parallel-tempering-md,sugita-kitao-okamoto:jcp:2000:hamiltonian-exchange,fukunishi-watanabe-takada:jcp:2002:hamiltonian-exchange,jang-shin-pak:prl:2003:hamiltonian-exchange,kwak-hansmann:prl:2005:hamiltonian-exchange}.
One such scheme involves attempting to exchange either the set of state index pairs $\{(1,2), (3,4), \ldots\}$ or $\{(2,3), (4,5), \ldots\}$, chosen with equal probability~\cite{footnote5}.

Each state index pair $(i,j)$ exchange attempt is attempted independently, with the exchange of states $i$ and $j$ associated with configurations $x_i$ and $x_j$, respectively, accepted with probability
\begin{eqnarray}
P_\mathrm{accept}(x_i, i, x_j, j) &=& \min\left\{ 1, \frac{e^{-[u_i(x_j)+u_j(x_i)]}}{e^{-[u_i(x_i) + u_j(x_j)]}}\right\}
\label{eq:metropolis-replica}
\end{eqnarray}

\subsubsection{Independence sampling}
\label{section:algorithms:replica-exchange:gibbs-sampling}

Independence sampling in replica exchange would consist of generating an uncorrelated, independent sample from $\pi(S|X)$.
The most straightforward scheme for doing so would require compiling a list of all possible $K!$ permutations of $S$, evaluating the unnormalized probability $\exp\left[-\sum_i u_{s_i}(x_i)\right]$ for each, normalizing by their sum, and then selecting a permutation $S'$ according to this normalized probability.
Even if the entire $K \times K$ matrix $\bfm{U} \equiv (u_{ij})$ with $u_{ij} \equiv u_i(x_j)$ is precomputed, the cost of this sampling scheme becomes impractical even for modestly large $K$.

Instead, we note that an \emph{effectively} uncorrelated sample from $\pi(S|X)$
can be generated by running an MCMC sampler scheme for a short time with trivial or small additional computational expense.
For each step of the MCMC sampler, we pick a pair of state indices $(i,j)$, with $i \ne j$, uniformly from the set $\{1, \ldots, K\}$.
The state pair associated with the configurations $x_i$ and $x_j$ are swapped with the same replica exchange Metropolis-like criteria shown in Eq.~\ref{eq:metropolis-replica},
with the labels of the states updated after each swap.
If we precompute the matrix $\bfm{U}$, then these updates are extremely inexpensive, and many Monte Carlo update steps of the state permutation vector $S$ can be taken to decorrelate from the previous sample for a fixed set of configurations $X$, effectively generating an uncorrelated sample $S' \sim P(S | X)$.  

In the case where all $u_{ij}$ are equal, then the number of swaps required is $K \ln K$---a well-known result due to 
\citet{aldous-diaconis:1986:amer-math-monthly:shuffling}.  
Empirically, we have found that swapping $K^3$ to $K^5$ times each state update iteration appears to be sufficient for the molecular cases examined in this paper and in our own work without consuming a significant portion of the replica exchange iteration time, but further experimentation may be required for some systems.  
Instead of performing random pair selections, we could also apply multiple passes of the neighbor exchange algorithm (Section~\ref{section:algorithms:replica-exchange:neighbor-exchange}). 
We note that complete mixing in state space is not a requirement for validity of the algorithm, but increasing the number of swaps will lead to increased space sampling until the limit of independent sampling is reached.

The method of multiple consecutive state swaps between configuration sampling is not entirely novel---we have heard several anecdotal examples of people experimenting with multiple consecutive state swaps, with sparse mentions in the literature~\cite{pitera_understanding_2003,martin-mayor:prb:2009:replica-exchange}.
However, we believe this is the first study to characterize the theory and properties of this particular modification of standard replica exchange.

For parallel tempering, in which only the inverse temperature $\beta_k$ varies with state index $k$, computation of $\bfm{U}$ is trivial if the potential energies of all $K$ states are known.
On the other hand, computation of all $u_i(x_j)$ for all $i,j = 1,\ldots,K$ may be time-consuming if the potential energy must be recomputed for each state, such as in an alchemical simulation.
If the Bennett acceptance ratio (BAR)~\cite{bennett:jcp:1976:fe-estimate} or the improved multistate version MBAR~\cite{shirts-chodera:jcp:2008:mbar} are used to analyze data generated during the simulation, however, all such energies are required anyway, and so no extra work is needed if the state update interval matches the interval at which energies are written to disk.
Alternatively, if the number of Monte Carlo or molecular dynamics time steps in between each state update is large compared to $K$, the overall impact on simulation time of the need to compute $\bfm{U}$ will be minimal.

\subsubsection{Other schemes}

The list of replica exchange methods above is by no means exhaustive---other schemes can be used for updating the state index $k$, provided they sample from the space of permutations $\pi(S | X)$ in a way that preserves the conditional distribution.
For example, it may be efficient for a node of a parallel computer to perform many exchanges only among replicas held in local memory, and to attempt few exchanges between nodes due to network constraints.
Compositions of different schemes are again also permitted.

%%%%%%%%%%%%%%%%%%%%%%%%%%%%%%%%%%%%%%%%%%%%%%%%%%%%%%%%%%%%%

\subsection{Metrics of efficiency}
\label{section:algorithms:metrics-of-efficiency}

There is currently no universally accepted metric for assessing sampling efficiency in molecular simulation, and thus it is difficult to quantify exactly how much our proposed algorithmic modifications improve sampling efficiency.
In the end, efficient algorithms will decrease the computational effort to achieve an estimate of the desired statistical precision for the expectations or free energy differences of interest.
Unfortunately, this can depend strongly on property of interest, the thermodynamic states that are being sampled, and the dynamics of the system studied.
While there exist metrics that describe the \emph{worst case} convergence behavior by approximating the slowest eigenvalue of the Markov chain~\cite{garren-smith:bernoulli:2000:estimating-second-eigenvalue,zuckerman:jctc:2010:effective-sample-size}, the worst case behavior can often differ from practical behavior by orders of magnitude~\cite{diaconis:contemporary-math:1992:metropolis-running-time}.
Here, we make use of a few metrics that will help us understand the time scale of these correlations in sampling under practical conditions.

Complex systems often get stuck in metastable states in configuration space with residence times a substantial fraction of the total available simulation time.   This dynamical behavior hinders the sampling of uncorrelated configurations by molecular dynamics simulation or Metropolis Monte Carlo schemes~\cite{schuette:j-comput-phys:1999:conformational-dynamics,schuette:2002:metastable-states}.
Systems can remain stuck in these metastable traps even as a replica in an expanded ensemble or replica exchange simulation travels through multiple thermodynamic states~\cite{huang-bowman-bacallado-pande:pnas:2009:adaptive-seeding}, either because the trap exists in multiple thermodynamic states or because the system does not have enough time to escape the trap before returning to states where the trap exists.
While approaches for detecting and characterizing the kinetics of these metastable states exist~\cite{defulhard-weber:lin-alg-appl:2005:pcca+,huang-bowman-bacallado-pande:pnas:2009:adaptive-seeding}, the combination of conformation space discretization error and statistical error makes the use of these approaches to compute relaxation times in configuration space not ideal for our purposes.

Here, we instead consider three simple statistics of the observed state index of each replica trajectory as surrogates to assess the improvements in overall efficiency of sampling.
Instead of considering the full expanded ensemble simulation trajectory $\{(x^{(0)},k^{(0)}), (x^{(1)},k^{(1)}), \ldots \}$ or the replica exchange simulation trajectory $\{(X^{(0)},S^{(0)}), (X^{(1)},S^{(1)}), \ldots \}$, we consider the trajectory of individual replicas projected onto the sequence of thermodynamic state indices $\bfm{s} \equiv \{s_0, s_1, \ldots\}$ visited during the simulation.
In long replica exchange simulations, each replica executes an equivalent random walk, and statistics can be pooled~\cite{chodera:jctc:2007:parallel-tempering-wham}.
If significant metastabilities in configuration space exist, we hypothesize that these configurational states will have different typical reduced potential $u(x)$ distributions, and therefore induce metastabilities in the state index trajectory $\bfm{s}$ as well that will be detectable by the methods described below. Each of the measures provides a different way to interpret the mixing of the simulation in state space; we will refer to all of them in the rest of the paper  as ``mixing times.''

\subsubsection{Relaxation time from empirical state transition matrix, $\tau_2$}
\label{section:applications:metrics-of-efficiency:second-eigenvalue}

One way to characterize how rapidly the simulation is mixing in state space is to examine the \emph{empirical transition matrix} among states, the $K \times K$ row-stochastic matrix $\bfm{T}$.
An individual element of this matrix, $T_{ij}$, is the probability that an expanded ensembles or replica exchange walker currently in state $i$ will be found in state $j$ the next iteration.
From a given expanded ensemble or replica exchange simulation, we can estimate $\bfm{T}$ by examining the expanded ensemble trajectory history or pooled statistics from individual replicas,
\begin{eqnarray}
T_{ij} &\approx& \frac{N_{ij} + N_{ji}}{\sum\limits_{k=1}^K [N_{ik} + N_{ki}]} \label{equation:empirical-state-transition-matrix}
\end{eqnarray}
where $N_{ij}$ is the number of times the replica is observed to be in state $k$ one update interval after being in state $i$.
To obtain a transition matrix $\bfm{T}$ with purely real eigenvalues, we have assumed both forward and time-reversed transitions in state indexes are equally probable, which is true in the limit of infinite time for all methods described in this paper. 
To assess how quickly the simulation is transitioning between different thermodynamic states, we compute the eigenvalues $\{\mu_1, \mu_2, \ldots, \mu_K\}$ of $\bfm{T}$ and sort them in descending order, such that that $1 = \mu_1 \ge \mu_2 \ge \cdots \ge \mu_K$.
If $\mu_2 = 1$, the Markov chain is \emph{decomposable}, meaning that two more subsets of the thermodynamic states exist where \emph{no} transitions have been observed between these sets, a clear indicator of very poor mixing in the simulation.  
In this case, the thermodynamic states characterized by $\{\lambda_1, \ldots, \lambda_K\}$ should be adjusted, or additional thermodynamic states inserted to enhance overlap in problematic regions.
Several schemes for optimizing the choice of these state vectors exist~\cite{kofke:2002:jcp:acceptance-probability,katzberger-trebst-huse-troyer:j-stat-mech:2006:feedback-optimized-parallel-tempering,trebst-troyer-hansmann:jcp:2006:optimized-replica-selection,nadler-hansmann:pre:2007:generalized-ensemble,gront-kolinski:j-phys-condens-matter:2007:optimized-replica-selection,park-pande:pre:2007:choosing-weights-simulated-tempering,shenfeld-xu:pre:2009:thermodynamic-length}, but are beyond the scope of this work to discuss here.

If the second-largest eigenvalue $\mu_2$ is such that $0 < \mu_2 < 1$ we can estimate a corresponding {\em relaxation time} $\tau_2$ as
\begin{eqnarray}
\tau_2 &=& \frac{\tau}{1 - \mu_2}
\end{eqnarray}
where $\tau$ is the effective time between exchange attempts.
$\tau_2$ then provides an estimate of the total simulation time required for the autocorrelation function in the state index $k^{(n)}$ of a replica at iteration $n$ of the simulation to decay to $1/e$ of the initial value. This estimate holds if the time scale of decorrelation in the configurational coordinate $x$ is fast compared to the decorrelation of the state index $k$; that is, if essentially uncorrelated samples could be drawn from $\pi(x | k)$ for each update of $x^{(n+1)} | k^{(n)}$.
Because configuration updates for useful molecular problems generally have long correlation times, this $\tau_2$ time represents a lower bound on the observed correlation time for both the state index $k^{(n)}$ and the configuration $x^{(n)}$.

\subsubsection{Correlation time of the replica state index, $\tau_\mathrm{ac}$}
\label{section:applications:metrics-of-efficiency:integrated-autocorrelation-time}

As a more realistic estimate of how quickly correlations in the state index $k^{(n)}$ decay in a replica trajectory, we also directly compute the correlation time of the state index history using the efficiency computation scheme described in Section 5.2 of \cite{chodera:jctc:2007:parallel-tempering-wham}, where $\tau_\mathrm{ac}$ is equal to the integrated area under the autocorrelation function.
For replica exchange simulations, where all replicas execute an equivalent walk in state space, the unnormalized autocorrelation functions were averaged over all replicas before computing the autocorrelation time by integrating the area under the autocorrelation function.
This time, $\tau_\mathrm{ac}$, gives a practical estimate of how much simulation time must elapse for correlations in the state index to fall to $1/e$. The {\em statistical inefficiency} is the number of samples required to collect each uncorrelated sample, and can be estimated for a Markovian process by $2\tau_{ac}+1$, with $\tau_{ac}$ in units of time between samples.

\subsubsection{Average end-to-end transit time of the replica state index, $\tau_\mathrm{end}$}
\label{section:applications:metrics-of-efficiency:transit-time}

As an additional estimate of practical efficiency, we measure the average end-to-end transition time for the state index, $\tau_\mathrm{end}$. 
This is the average of the time elapsed between the first visit of the state index $k^{(n)}$ to one end point ($k=1$ or $k=K$) after visiting the opposite end point ($k=K$ or $k=1$, respectively).
This metric of efficiency, or the related ``round-trip'' time, has seen common use in diagnosing efficiency for simulated-tempering and replica exchange simulations~\cite{trebst-troyer-hansmann:jcp:2006:optimized-replica-selection,nadler-hansmann:pre:2007:optimized-replica-exchange-moves,escobedo-martinez-veracoechea:jcp:2008:optimization-of-expanded-ensemble,denschlag-lingenheil-tavan:cpl:2009:optimal-temperature-ladders}.

%%%%%%%%%%%%%%%%%%%%%%%%%%%%%%%%%%%%%%%%%%%%%%%%%%%%%%%%%%%%%
\section{Model Illustration}
\label{section:illustration}

\begin{figure}[tbp]
\noindent
\resizebox{\columnwidth}{!}{\includegraphics{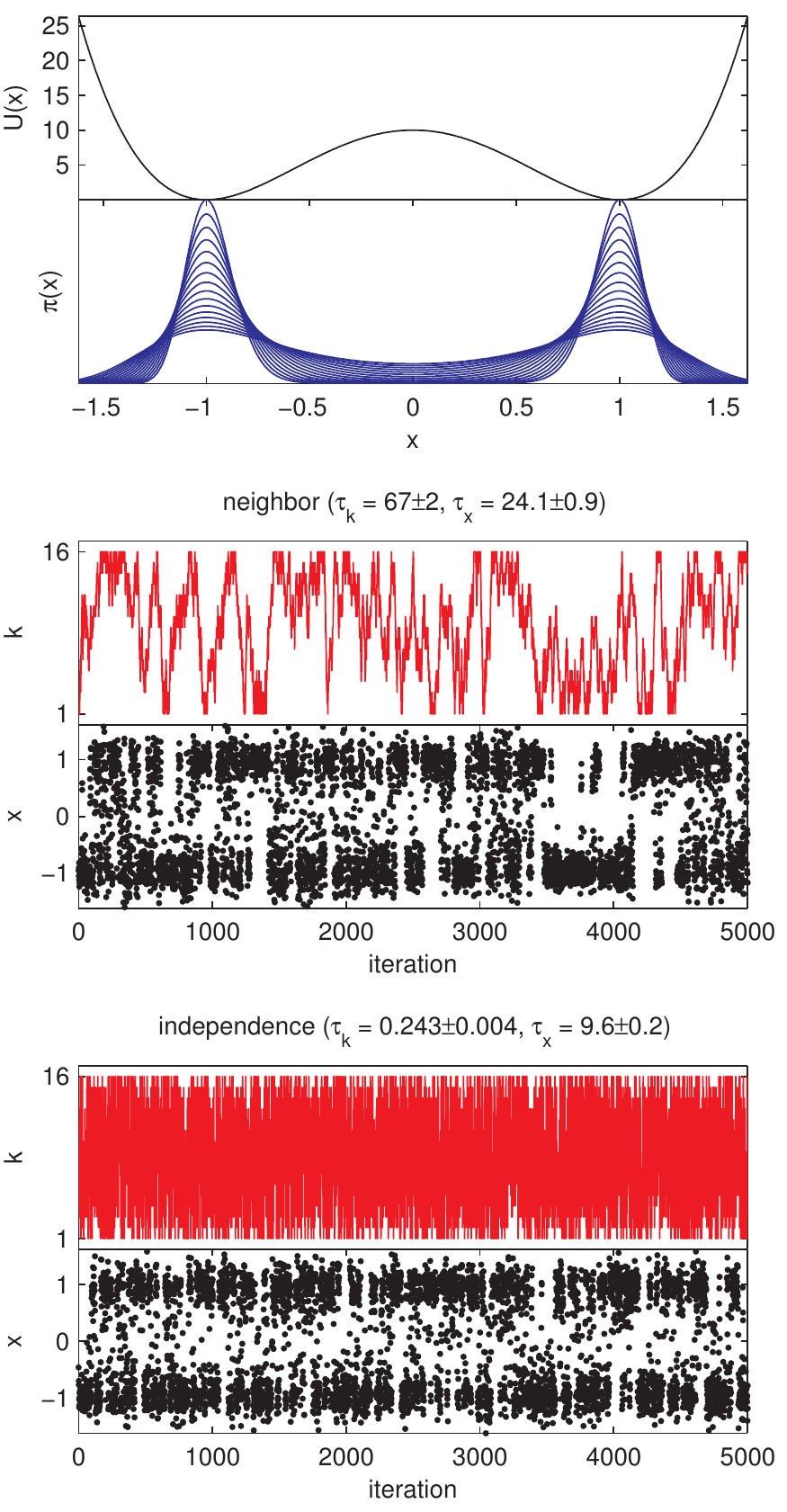}}
\caption{{\bf Simulated tempering for a one-dimensional model.}
\emph{Top panel:} Potential energy $U(x)$ and stationary probabilities $\pi(x)$ for one-dimensional two-well model potential at 16 temperatures spanning $k_B T = 1$, where barrier crossing is hindered, to $k_B T = 10$, where barrier crossing is rapid.
\emph{Middle panel:} Temperature index, $k$, and position, $x$, histories for a simulated tempering simulation where neighbor swap in temperature are attempted each iteration.
\emph{Bottom panel:} Temperature index, $k$, and position, $x$, histories for a simulated tempering simulation where independence sampling of the temperature index is performed each iteration.
Only the first 5 000 iterations are shown, though simulations of $10^6$ iterations were conducted to estimate the correlation times $\tau_k$ and $\tau_x$ printed above each panel, shown in number of iterations required to produce an effectively uncorrelated sample in either $k$ or $x$, respectively.
Statistical uncertainties shown represent one standard error of the mean.
\label{figure:model-example-trajectory}}
\end{figure}

\begin{figure}[tbp]
\noindent
\resizebox{\columnwidth}{!}{\includegraphics{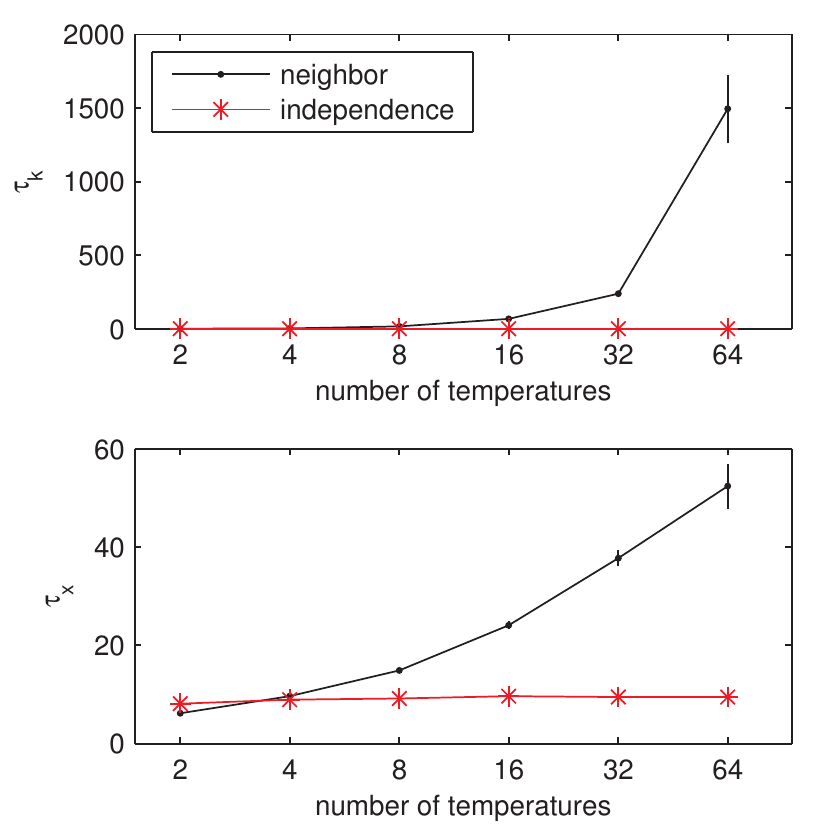}}
\caption{{\bf Autocorrelation times as function of number of temperatures for one-dimensional model.}
The integrated autocorrelation time $\tau_k$ for state index $k$ (\emph{top}) and $\tau_x$ for position $x$ (\emph{bottom}) as a function of the number of exponentially-spaced temperatures spanning the range $k_B T \in [1,10]$.
The correlation times for neighbor swap (\emph{black points}) and independence sampling updates (\emph{red stars}) are shown for each.
Error bars represent one standard error of the mean.
\label{figure:model-g-vs-ntemps}}
\end{figure}

To illustrate the motivation behind the idea that speeding up sampling in one coordinate---the state index or permutation---will enhance sampling of the overall Markov chain of $(x,k)$ or $(X,P)$, we consider a simulated tempering simulation in a one-dimensional model potential,
\begin{eqnarray}
U(x) &=& 10 (x-1)^2 (x+1)^2 .
\end{eqnarray}
shown in the top panel of Figure~\ref{figure:model-example-trajectory}, along with the corresponding stationary distribution $\pi(x)$ at several temperatures from $k_B T = 1$ to $k_B T = 10$.
To simplify our illustration, we directly numerically compute the log-weight factors
\begin{eqnarray}
g_k &=& - \ln \int_{-\infty}^{+\infty} dx \, e^{-\beta_k U(x)}
\end{eqnarray}
so that the simulation has an equal probability to be in each of the $K$ states.

The $K$ inverse temperatures $\beta_k$ that can be visited during the simulated tempering simulation are chosen to be geometrically spaced,
\begin{eqnarray}
\beta_k &=& 10^{-(k-1)/(K-1)} \:\: \mathrm{for} \:\: k = 1,\ldots,K \label{equation:geometric-temperature-spacing}
\end{eqnarray}

Each iteration of the simulation consists of an update of the temperature index $k$ using either neighbor exchange (Section~\ref{section:algorithms:expanded-ensemble:neighbor-exchange}) or independence sampling updates (Section~\ref{section:algorithms:expanded-ensemble:gibbs-sampling}), followed by 100 steps of Metropolis Monte Carlo~\cite{metropolis:jcp:1953:metropolis-monte-carlo,hastings:biometrika:1970:metropolis-hastings} using a Gaussian proposal with zero mean and standard deviation of 0.1 in the $x$-coordinate.
Simulations are initiated from $(x_0,k_0) = (-1,1)$.

Illustrative trajectories for $K = 16$ are shown in the second and third panels of Figure~\ref{figure:model-example-trajectory}, along with the correlation times $\tau_k$ and $\tau_x$ computed for the temperature index $k$ and the configurational coordinate $x$, respectively, from a long trajectory of $10^6$ iterations.
Independence sampling in state space $k$ greatly reduces the correlation time, and hence statistical inefficiency, in $k$ compared to neighbor sampling. Importantly, because $k$ and $x$ are coupled, we clearly see that increasing the mixing in the index $k$ also substantially reduces the correlation time in the configurational coordinate $x$.  We find that $\tau_x = 9.6 \pm 0.2$ for independence sampling, compared to $24.1 \pm 0.9$ for neighbor moves.

Figure~\ref{figure:model-g-vs-ntemps} compares the correlation times for $k$ and $x$ estimated from simulations of length $10^6$ for different numbers of temperatures spanning the same range of $k_B T \in [1,10]$, with temperatures again geometrically spaced according to Eq.~\ref{equation:geometric-temperature-spacing}.
As the number of temperatures spanning this range increases, the correlation time in the temperature coordinate $k$ increases, as one would expect for a random walk on domains of increasing size.
Notably, increasing the number of temperatures \emph{also} has the effect of increasing the correlation time of the configuration coordinate $x$.
When independence sampling is used to update the temperature index $k$ instead, the mixing time in $k$ is greatly reduced, and both correlation times $\tau_k$ and $\tau_x$ remain small even as the number of temperatures is increased.

%%%%%%%%%%%%%%%%%%%%%%%%%%%%%%%%%%%%%%%%%%%%%%%%%%%%%%%%%%%%%
% APPLICATIONS
%%%%%%%%%%%%%%%%%%%%%%%%%%%%%%%%%%%%%%%%%%%%%%%%%%%%%%%%%%%%%

\section{Applications}
\label{section:applications}

To demonstrate that the simple state update modifications we describe in Section~\ref{section:algorithms} lead to real efficiency improvements in practical simulation problems, we consider three typical simulation problems: An alchemical expanded ensemble simulation of united atom (UA) methane in water to compute the free energy of transfer from gas to water; a parallel tempering simulation of terminally-blocked alanine dipeptide in implicit solvent; and a two-dimensional replica exchange umbrella sampling simulation of alanine dipeptide in implicit solvent to compute the potential of mean force.   These systems are small compared to modern applications of biophysical and biochemical interest.  However, they are realistic enough to demonstrate the fundamental issues in multiensemble simulations, but still sufficiently tractable that a large quantity of data can be collected to prove that the differences in efficiency of our proposed mixing schemes are highly significant.

%%%%%%%%%%%%%%%%%%%%%%%%%%%%%
\subsection{Expanded ensemble alchemical simulations of Lennard-Jones spheres in water}

\subsubsection{United atom methane}

\begin{table*}[tbp]
\begin{tabular}{lccccccccc}
\hline
& \multicolumn{4}{c}{mixing times (ps)} && \multicolumn{4}{c}{relative speedup} \\ \cline{2-5}  \cline{7-10}
& $\tau_2$ & $\tau_\mathrm{ac}$ & $\tau_\mathrm{end}$ & $\tau_\mathrm{N}$ && $\tau_2$ & $\tau_\mathrm{ac}$ & $\tau_\mathrm{end}$ & $\tau_\mathrm{N}$ \\
\hline
\multicolumn{10}{l}{\bf 1 state move attempted every 0.1 ps} \\
\hline
neighbor exchange &   1.693 $\pm$ 0.008 &     6.7 $\pm$   0.4 &    11.9 $\pm$   0.2 &     5.9 $\pm$   0.4 && 1.0      & 1.0      & 1.0      & 1.0      \\
independence sampling \hspace{0.1in} &   0.771 $\pm$ 0.004 &     6.2 $\pm$   0.2 &     7.2 $\pm$   0.1 &     5.4 $\pm$   0.2 &&  2.20 $\pm$   0.02&  1.08 $\pm$   0.08&  1.65 $\pm$   0.04&  1.10 $\pm$   0.09\\
Metropolized indep. &   0.645 $\pm$ 0.003 &     4.6 $\pm$   0.2 &     6.6 $\pm$   0.1 &     4.4 $\pm$   0.2 &&  2.62 $\pm$   0.02&   1.5 $\pm$    0.1&  1.81 $\pm$   0.04&   1.3 $\pm$    0.1 \\
\hline
\multicolumn{10}{l}{\bf 1 000 state moves attempted every 0.1 ps} \\
\hline
neighbor exchange &   0.764 $\pm$ 0.006 &     4.9 $\pm$   0.3 &     7.2 $\pm$   0.1 &     4.6 $\pm$   0.3 && 1.0      & 1.0      & 1.0      & 1.0      \\
independence sampling \hspace{0.1in} &   0.769 $\pm$ 0.005 &     4.8 $\pm$   0.2 &     7.2 $\pm$   0.1 &     4.5 $\pm$   0.2 &&  0.99 $\pm$   0.01&  1.01 $\pm$   0.08&  1.00 $\pm$   0.02&  1.01 $\pm$   0.07\\
Metropolized indep. &   0.774 $\pm$ 0.005 &     5.0 $\pm$   0.2 &     7.5 $\pm$   0.1 &     4.7 $\pm$   0.2 &&  0.99 $\pm$   0.01&  0.98 $\pm$   0.08&  0.96 $\pm$   0.02&  0.98 $\pm$   0.07\\
\hline
\multicolumn{10}{l}{\bf 1 state move attempted every 5 ps} \\
\hline
neighbor exchange &    85.8 $\pm$   2.3 &   177.7 $\pm$  17.6 &   330.0 $\pm$  16.1 &   105.3 $\pm$  12.1 && 1.0      & 1.0      & 1.0      & 1.0      \\ 
independence  sampling \hspace{0.1in} &    39.0 $\pm$   0.9 &    69.2 $\pm$   6.1 &   141.1 $\pm$   4.7 &    49.1 $\pm$   3.8 &&  2.20 $\pm$   0.08&   2.6 $\pm$    0.3&   2.3 $\pm$    0.1&   2.1 $\pm$    0.3\\ 
Metropolized indep. &    31.8 $\pm$   0.4 &    51.4 $\pm$   1.9 &   115.7 $\pm$   3.4 &    37.4 $\pm$   1.4 &&  2.70 $\pm$   0.08&   3.5 $\pm$    0.4&   2.9 $\pm$    0.2&   2.8 $\pm$    0.3\\ 
\hline
\end{tabular}
\caption{{\bf Efficiency measures for expanded ensemble alchemical simulation of united atom methane in water.}
Times measuring mixing in state space are: 
$\tau_2$, estimated from second eigenvalue of the empirical state transition matrix; 
$\tau_\mathrm{ac}$, estimated from autocorrelation function of the alchemical state index; 
$\tau_\mathrm{end}$, estimated average end-to-end transit time for the alchemical state index.
$\tau_\mathrm{N}$ is a structural parameter, the autocorrelation function of the number of TIP3P oxygen molecules 
within 0.3 nm (87.5\% of the Lennard-Jones $\sigma_{ij}$ (0.3428 nm)) of the center of the united atom methane particle. 
The \emph{relative speedup} in sampling efficiency is given relative to the standard neighbor exchange scheme.
\label{table:expanded-ensemble-methane}
}
\end{table*}

We first compare different types of Gibbs sampling state space updates in an expanded ensemble alchemical simulation of the kind commonly used to compute the free energy of hydration of small molecules~\cite{fenwick-escobedo:jcp:2003:replica-exchange-expanded-ensembles,escobedo-martinez-veracoechea:jcp:2008:optimization-of-expanded-ensemble}.
If the state mixing schemes proposed here lead to more efficient sampling among alchemical states, a larger number of effectively uncorrelated samples will be generated for a simulation of a given duration, and thus require less computation effort to reach the desired degree of statistical precision.  

An OPLS-UA united atom methane particle ($\sigma = 0.373$ nm, $\epsilon=1.230096$ kJ/mol) was solvated in a cubic simulation cell containing 893 TIP3P~\cite{jorgensen:jcp:1983:tip3p} waters. 
For all simulations, a modified version of {\tt GROMACS} 4.5.2~\cite{gromacs4} was used~\cite{footnote6}.
A velocity Verlet integrator~\cite{swope:jcp:1982:velocity-verlet} was used to propagate dynamics with a timestep of 2 fs.  
A Nos\'{e}-Hoover chain of length 10~\cite{martyna_nose-hoover_1992} and time constant $\tau_T = 10.0$ ps was used to thermostat the system to 298 K.  A measure-preserving barostat was used according to Tuckerman et al.~\cite{yu_measure-preserving_2010,tuckerman_liouville-operator_2006} to maintain the average system pressure at 1 atm, with $\tau_p = 10.0$ ps and compressibility $4.5\times10^{-5}$ bar$^{-1}$. 
Rigid geometry was maintained for all waters using the analytical SETTLE scheme~\cite{kollman:1992:j-comput-chem:settle}. 
A neighborlist and PME cutoff of 0.9 nm were used, with a PME order of 6, spacing of 0.1 nm and a relative tolerance of $10^{-6}$ at the cutoff.  
The Lennard-Jones potential was switched off, with the switch beginning at 0.85 nm and terminating at the cutoff of 0.9 nm.
An analytical dispersion correction was applied beyond the Lennard-Jones cutoff to correct the energy and pressure computation~\cite{shirts_accurate_2007}. 
The neighborlist was updated every 10 steps.

A set of $K = 6$ alchemically-modified thermodynamic states were used in which the Lennard-Jones interactions between the methane and solvent were eliminated using a soft-core Lennard-Jones potential~\cite{shirts_solvation_2005},
\begin{eqnarray}
 U_{ij}(r;\lambda) &=& 4\epsilon_{ij}\lambda \, f(r;\lambda) [1 - f(r;\lambda)] \nonumber \\
f(r;\lambda) &\equiv& [\alpha(1-\lambda) + (r/\sigma_{ij})^6]^{-1}
\end{eqnarray}
with values of the alchemical coupling parameter $\lambda_k$ chosen to be $\{0.0,0.3,0.6,0.7,0.8,1.0\}$.

To simplify our analysis of efficiency, we fix the log-weights $g_k$ to ``perfect weights,'' where all states are visited with equal probability.
This also decouples the issue of efficiency of state updates with efficiency of different weight update schemes, of which many have been proposed~\cite{lyubartsev:jcp:1992:expanded-ensembles,marinari-parisi:europhys-lett:1992:simulated-tempering,wang-landau:prl:2001:wang-landau,park-ensign-pande:pre:2006:bayesian-weight-update,park-pande:pre:2007:choosing-weights-simulated-tempering,li-fajer-yang:jcp:2007:simulated-scaling,chelli:jctc:2010:optimal-weights-expanded-ensembles}.
The ``perfect'' log-weights were estimated for this system as follows: 
A 1~ns expanded ensemble simulation using independence sampling was run, with weights $g_k$ initialized to zero, then adjusted using a Wang-Landau scheme~\cite{li-fajer-yang:jcp:2007:simulated-scaling}, until occupancy of each state was roughly even to within statistical noise.    
With these approximate weights, a 2~ns expanded ensemble simulation using independence sampling with fixed weights was run, and the free energy of each state was estimated using MBAR~\cite{shirts-chodera:jcp:2008:mbar}.  
The log-weights $g_k$ were set to these estimated free energies, which were $\{0.0, 0.32, -0.46, -1.67, -2.83, -3.66\}$, in units of $k_{B}T$.  
Simulations using these weights deviated by an average of 5\% from flat histogram occupancy in states, with an average maximum deviation over all simulations of less than 10\%.  

The state update procedure was carried out either every 0.1~ps (frequent update) or 5~ps (infrequent update), in order to test the effect of state updates that were much faster than, or on the order of, the conformational correlation times of molecular dynamics, as water orientational correlation times are a few picoseconds~\cite{shirts_solvation_2003}.
Production simulations with fixed log-weights were run with for 25~ns (250 000 state updates), for frequent updates, or 100~ns (20 000 state updates), for infrequent updates.  
Three types of state moves were attempted: 
(1) neighbor exchange moves (described in Section~\ref{section:algorithms:expanded-ensemble:metropolized-gibbs}),
(2) independence sampling (Section~\ref{section:algorithms:expanded-ensemble:gibbs-sampling}), 
and (3) Metropolized independence sampling (Section~\ref{section:algorithms:expanded-ensemble:metropolized-gibbs}).
In the case of frequent updates, we additionally performed 1 000 trials of the state update every 0.1~ps, instead of a single update, before returning to coordinate update moves with molecular dynamics.

Statistics of the observed replica trajectories are shown in Table~\ref{table:expanded-ensemble-methane}.  
All three mixing efficiency measures of the state index trajectories described in Section~\ref{section:algorithms:metrics-of-efficiency} were computed: relaxation time of the empirical state transition matrix ($\tau_2$), autocorrelation of the state function ($\tau_{ac}$), and average end-to-end distance ($\tau_{\mathrm{end}}$).

We additionally look at a measure of correlation in the coordinate direction.  
For each configuration, we examine the number of O atoms of the water molecules $N$ that are found in the interior of the united atom methane, set to be 0.3 nm (or 87.5\% of the Lennard-Jones $\sigma_{ij}=0.3428$ nm) from the center.  
We then compute the autocorrelation function of $\tau_{N}$ of this variable, which is affected both by the dynamics of the state and the dynamical
response of the system to changes in state.  
Uncertainties in these time autocorrelation functions are computed by subdividing the trajectories into $N_S=10$ subtrajectories, computing the standard error, and then dividing by $\sqrt{N_S}$ to obtain standard error of the $N_S\times$ longer trajectory.  
Uncertainties changed by less than 5\% when computed with $N_S=20$ for frequent update simulations, and less than 10\% for infrequent update simulations.

The relaxation time $\tau_2$ estimated from the second eigenvalue of the empirical state transition matrix (Section~\ref{section:applications:metrics-of-efficiency:second-eigenvalue}) does appear to provide a lower bound for the other estimated mixing times.  
For the infrequent state updates, it is only about 25\% smaller than $\tau_{N}$.  
This suggests that when transition times in state space are of the same order of magnitude as conformational rearrangements $\tau_2$ is not only a lower bound, but is characteristic of sampling through the joint state-configuration space.
We additionally note that mixing time $\tau_2$ is empirically exactly proportional to the update frequency; the mixing times for the infrequent update state are exactly (5 ps/0.1 ps) = 50 times longer than the frequent state mixing times, a direct consequence of the fact that the probability of successful state transitions is directly proportional to the rate of attempted transitions.

For both the frequent and infrequent state updates, independence sampling and Metropolized independence sampling yield a clear, statistically significant speedup by all sampling metrics.  
This speedup is accentuated for infrequent updates.  
For frequent updates, the speedup is between 1.3 and 2.6 for Metropolized independence sampling, while for infrequent updates, it ranges between 2.7 and 3.5, as seen in Table~\ref{table:expanded-ensemble-methane}. 
As expected, attempting many state updates in a row (1 000 state moves) using any of the state update schemes effectively recapitulates the independence sampling scheme.
Repeated application of any method that obeys the balance condition will eventually converge to the same independent sampling distribution.  If state updates are relatively inexpensive, then any state update scheme that ensures the correct distribution is sampled can be iterated many times, effectively resulting in an independence sampling scheme.
Interestingly, this means that Metropolized independence sampling becomes worse when repeated several times, as it eventually turns into simple independence sampling.

Although the acceleration of independence sampling over neighbor exchange is more dramatic with longer intervals between state updates, more frequent state updates appear to always be better than less frequent updates. For example, neighbor exchange with more frequent updates achieves shorter correlation times that either independence sampling scheme for infrequent updates.  
Increased sampling frequency in state space seems to be a good idea.~\cite{sindhikara-meng-roitberg:jcp:2008:exchange-frequency,sindhikara-emerson-roitberg:jctc:2010:exchange-often-and-properly}
It is possible that there are conditions where this conclusion might not be true; collective moves like long molecular dynamics trajectories of polymers might become disrupted by too frequent changes in state space.  Additional study is required to understand this phenomena.
We finally note that for this particular system, Metropolized independence sampling is slightly but clearly better than independence sampling in all sampling measures, providing a strong incentive to use Metropolized independence sampling where convenient.

\subsubsection{Larger Lennard-Jones spheres}

\begin{table*}[tbp]
\begin{tabular}{lccccccccc}
\hline
& \multicolumn{4}{c}{mixing times (ps)} && \multicolumn{4}{c}{relative speedup} \\ \cline{2-5}  \cline{7-10}
& $\tau_2$ & $\tau_\mathrm{ac}$ & $\tau_\mathrm{end}$ & $\tau_\mathrm{N}$ && $\tau_2$ & $\tau_\mathrm{ac}$ & $\tau_\mathrm{end}$ & $\tau_\mathrm{N}$ \\
\hline
\multicolumn{10}{l}{\bf 1 state move attempted every 0.1 ps} \\
\hline
neighbor exchange &    9.51 $\pm$  0.01 &    65.8 $\pm$   4.2 &   126.3 $\pm$   4.2 &    58.1 $\pm$   4.3 && 1.0      & 1.0      & 1.0      & 1.0      \\
independence sampling \hspace{0.1in}   &   2.586 $\pm$ 0.009 &    42.9 $\pm$   2.4 &    88.4 $\pm$   2.7 &    41.5 $\pm$   2.0 &&  3.68 $\pm$   0.01&   1.5 $\pm$    0.1&  1.43 $\pm$   0.06&   1.4 $\pm$    0.1\\
Metropolized indep. &   2.181 $\pm$ 0.006 &    48.6 $\pm$   4.0 &    88.3 $\pm$   3.0 &    46.7 $\pm$   3.4 &&  4.36 $\pm$   0.01&   1.4 $\pm$    0.1&  1.43 $\pm$   0.07&   1.2 $\pm$    0.1\\
\hline
\multicolumn{10}{l}{\bf 1 state move attempted every 1 ps} \\
\hline
neighbor exchange&    95.0 $\pm$   0.2 &   211.1 $\pm$  58.9 &   507.6 $\pm$  19.3 &   167.6 $\pm$  16.0 && 1.0      & 1.0      & 1.0      & 1.0      \\ 
independence sampling \hspace{0.1in} &    25.8 $\pm$   0.1 &    67.3 $\pm$   3.6 &   196.0 $\pm$   5.8 &    63.1 $\pm$   3.3 &&  3.69 $\pm$   0.02&   3.1 $\pm$    0.9&   2.6 $\pm$    0.1&   2.7 $\pm$    0.3\\ 
Metropolized indep. &    21.6 $\pm$   0.1 &    66.8 $\pm$   2.4 &   169.2 $\pm$   4.7 &    62.1 $\pm$   2.5 &&  4.40 $\pm$   0.02&   3.2 $\pm$    0.9&   3.0 $\pm$    0.1&   2.7 $\pm$    0.3\\ 
\hline
\end{tabular}
\caption{{\bf Efficiency measures for expanded ensemble alchemical simulation of large LJ sphere in water.}
Times measuring mixing in state space are: 
$\tau_2$, estimated from second eigenvalue of empirical state transition matrix; 
$\tau_\mathrm{ac}$, estimated from autocorrelation time of alchemical state index; 
$\tau_\mathrm{end}$, estimated average end-to-end transit time for alchemical state index.
$\tau_\mathrm{N}$ is a structural parameter, the autocorrelation function of the number of TIP3P oxygen molecules 
within 0.5 nm (85.3\% of the Lennard-Jones $\sigma_{ij}$ (0.5860 nm)) of the center of the large Lennard-Jones particle. 
The \emph{relative speedup} in sampling efficiency is given relative to the standard neighbor exchange scheme.
\label{table:expanded-ensemble-bigLJ}
}
\end{table*}

As united atom methane is much smaller than typical biomolecules of interest, we additionally examined an alchemical expanded ensemble simulation of a much larger Lennard-Jones sphere.  
In this case, the sphere has $\sigma_{ii} = 1.09$ nm and $\epsilon_{ii}=1.230096$ kJ/mol, again solvated in a cubic simulation cell containing 893 TIP3P~\cite{jorgensen:jcp:1983:tip3p} waters. 
These parameters result in a sphere-water $\sigma_{ij}=0.561$ nm, and therefore a particle $5.0$ times as large in volume as the UA methane sphere.  
Because of the larger volume of the solute, $K = 18$ alchemically-modified thermodynamic states were required, with $\lambda$ = [$0$, $0.15$, $0.3$, $0.45$, $0.55$, $0.6$, $0.64$, $0.66$, $0.68$, $0.70$, $0.72$, $0.75$, $0.78$, $0.81$, $0.84$, $0.87$, $0.90$, $1.0$]. 
All other simulation parameters (other than simulation length) were the same as the UA methane simulations.
Log-weights $g_k$ for the equilibrium expanded ensemble simulation were determined in the same manner as for united atom methane, except that a 15~ns simulation was used to generate the data for MBAR, yielding weights $g_k$ = $\{0.0$, $1.74$, $2.96$, $3.39$, $2.84$, $2.01$, $0.73$, $-0.34$, $-1.75$, $-3.35$, $-4.96$, $-7.19$, $-9.11$, $-10.70$, $-11.98$, $-12.98$, $-13.72$, $-14.65\}$.
Frequent state updates were performed every 0.1~ps, but infrequent state moves were performed every 1~ps rather than 5~ps to obtain better statistics for the larger molecule.  
The production expanded ensemble simulations were run for a total of 100 ns for frequent exchange, and 250 ns for infrequent exchange.  The same three types of moves in state space were attempted as with UA methane.

Statistics of the observed replica trajectories are shown in Table~\ref{table:expanded-ensemble-bigLJ}.  
All three convergence rate diagnostics of the state index trajectories described in Section~\ref{section:algorithms:metrics-of-efficiency} were computed.
In general, the relaxation time estimated from the second eigenvalue of the empirical state transition matrix (Section~\ref{section:applications:metrics-of-efficiency:second-eigenvalue}) again provides a lower bound for the other computed relaxation times. For the infrequent sampling interval $\tau_2$ is of the same order of magnitude (2 to 5 times less) than the other sampling measures.  
Again, for both the frequent (0.1~ps) and infrequent (1~ps) state update intervals, independence sampling and Metropolized independence sampling yields a clear speedup over neighbor exchange.  The improvement in sampling efficiency appears to be valid for both small and large particles.

%%%%%%%%%%%%%%%%%%
\subsection{Parallel tempering simulations of terminally-blocked alanine peptide in implicit solvent}

\begin{table*}[tbp]
\begin{tabular}{lc|c|c||c|c|c|c}
\hline
& \multicolumn{3}{c||}{state mixing times (ps)} & \multicolumn{4}{c}{structural correlation times (ps)} \\
& $\tau_2$ & $\tau_\mathrm{ac}$ & $\tau_\mathrm{end}$ & $\tau_{\cos \phi}$ & $\tau_{\sin \phi}$ & $\tau_{\cos \psi}$ & $\tau_{\sin \psi}$\\
\hline
neighbor exchange & 91.8 $\pm$ 0.6 & 80 $\pm$ 2 & 360 $\pm$ 30 & 25 $\pm$ 2 & 110 $\pm$ 9 & 25 $\pm$ 2 & 66 $\pm$ 6 \\
independence sampling \hspace{0.1in}  & 2.62 $\pm$ 0.01 & 1.60 $\pm$ 0.06 & 28.7 $\pm$ 0.7 & 12.4 $\pm$ 0.5 & 8.7 $\pm$ 0.4 & 11.8 $\pm$ 0.6 & 9.1 $\pm$ 0.5 \\
\hline
\end{tabular}
\caption{{\bf Efficiency measures for parallel tempering simulation of alanine dipeptide in implicit solvent.}
Mixing times listed are: 
$\tau_2$, estimated from second eigenvalue of empirical state transition matrix; 
$\tau_\mathrm{ac}$, estimated from autocorrelation time of alchemical state index; 
$\tau_\mathrm{end}$, estimated average end-to-end transit time for alchemical state index.
Autocorrelation times of trigonometric functions of $\phi$ and $\psi$ torsion angles are listed as $\tau_{\cos \phi}, \tau_{\sin \phi}, \tau_{\cos \psi}, \tau_{\sin \psi}$.
The statistical error is given as one standard error of the mean.
\label{table:alanine-dipeptide-parallel-tempering}
}
\end{table*}

We next consider a parallel tempering simulation, a form of replica exchange in which the thermodynamic states differ only in inverse temperature $\beta_k$. A system containing terminally-blocked alanine (sequence Ace-Ala-Nme) was constructed using the {\tt LEaP} program~\cite{ambertools10-leap} from the {\tt AmberTools} 1.2 package with bugfixes 1--4 applied.
The Amber parm96 forcefield was used~\cite{AMBER-parm96} along with the Onufriev-Bashford-Case generalized Born-surface area (OBC GBSA) implicit solvent model (corresponding to model I of~\cite{onufriev-bashford-case:2004:proteins:obc-gbsa}, equivalent to {\tt igb=2} in Amber's {\tt sander} program and using the {\tt mbondi2} radii selected within {\tt LEaP}).

A custom Python code making use of the GPU-accelerated {\sc OpenMM} package~\cite{friedrichs:2009:j-comput-chem:openmm,eastman:2010:comp-sci-eng:openmm,eastman:2010:j-comput-chem:openmm} and the {\sc PyOpenMM} Python wrapper~\cite{pyopenmm} was used to conduct the simulations.
All forcefield terms are identical to those used in AMBER except for the surface area term, which was left as default in the OpenMM implementation through a GBSAOBCForce term. 
Parallel tempering simulations of 2 000 iterations were run, with dynamics propagated by 500 steps each iteration using a 2 fs timestep and the leapfrog Verlet integrator~\cite{verlet:1967:phys-rev:verlet-integrator-1,verlet:1967:phys-rev:verlet-integrator-2}.
Velocities were reassigned from the Maxwell-Boltzmann distribution each iteration.
The Python scripts for simulation and data analysis used here are available online at \url{http://simtk.org/home/gibbs}.

For the replica-mixing phase, the simulation employed either neighbor exchange (Section~\ref{section:algorithms:replica-exchange:neighbor-exchange}) or independence sampling (Section~\ref{section:algorithms:replica-exchange:gibbs-sampling}), with $K^3$ attempted swaps of replica pairs selected at random.
The efficiency was measured in several ways, shown in Table~\ref{table:alanine-dipeptide-parallel-tempering}.
In addition to the standard mixing metrics described in Section~\ref{section:algorithms:metrics-of-efficiency}, an estimate of the configurational relaxation times was also made; due to the circular nature of the torsional coordinates $\phi$ and $\psi$ known to be slow degrees of freedom for this system~\cite{chodera:mms:2006:long-time-dynamics}, we instead computed the autocorrelation times for $\sin\phi$, $\cos\phi$, $\sin\psi$, and $\cos\phi$.
All replicas were treated as equivalent, and their raw statistics (e.g.~autocorrelation functions before normalization) were averaged to produce these estimates. Statistical error was again estimated by blocking.

As expected, the various metrics indicate that the parallel tempering replicas mix in state space much more rapidly with independence sampling than when only neighbor exchanges are attempted.
The amount by which mixing is accelerated depends on the metric used to quantify this, but it is roughly one to two orders of magnitude.
The structural relaxation times also reflect a speedup, though much more modest than the acceleration in state space sampling---roughly a factor of two to ten, depending on the metric examined.

%%%%%%%%%%%%%%%%%%
\subsection{Two-dimensional replica exchange umbrella sampling of terminally-blocked alanine peptide in implicit solvent}

Finally, we consider a two-dimensional replica exchange umbrella sampling situation, commonly used to compute potentials of mean force along two coordinates of interest.
We again consider the alanine dipeptide in implicit solvent, and employ umbrella potentials to restrain the $\phi$ and $\psi$ torsions near reference values $(\phi^0_k, \psi^0_k)$ for $K = 101$ replicas spaced evenly on a $10 \times 10$ toroidal grid, with the inclusion of one replica without any bias potential for ease of post-simulation analysis.

Because harmonic constraints are not periodic, we employ periodic bias potential based on the von Mises circular normal distribution,
\begin{eqnarray}
U'_k(x) &\equiv& - \kappa \left[ \cos(\phi - \phi^0_k) + \cos(\psi - \psi^0_k) \right]
\end{eqnarray}
where $\kappa$ has units of energy.
For sufficiently large values of $\kappa$, this will localize the torsion angles in an approximately Gaussian distribution near the reference torsions $(\phi^0_k,\psi^0_k)$ with a standard deviation of $\sigma \equiv (\beta \kappa)^{1/2}$.

Here, we employ a $\kappa$ of $(2 \pi / 30)^{-2} \beta^{-1}$ so that neighboring bias potentials are separated by $3 \sigma$.
This was sufficient to localize sampling near the reference torsion values for most sterically unhindered regions.
The simulation was run at 300 K, using a 2 fs timestep with 5 ps between replica exchange attempts.
A total of 2 000 iterations were conducted, with each iteration consisting of mixing the replica state assignments via a state update phase, a new velocity assignment from the Maxwell-Boltzmann distribution, propagation of dynamics, and writing out the resulting configuration data.  The first 100 iterations were discarded as equilibration.

The same mixing schemes examined in the parallel tempering simulation were evaluated here, and the results of the efficiency metrics are summarized in Table~\ref{table:alanine-dipeptide-2d-umbrella-sampling}.
Note that the end-to-end time does not have a clear interpretation in terms of the average transit time between a maximum and minimum thermodynamic parameter here---it simply reflects the average time between exchanges between a particular localized umbrella and the unbiased state.

As in the parallel tempering case, we find that both mixing times in state space and the structural correlation times are reduced by use of Gibbs sampling, albeit to a lesser degree than in the parallel tempering case.
Here, state relaxation times are reduced by a factor of two to six, depending on the metric considered, while structural correlation times are reduced by a factor of four or five.

\begin{table*}[tbp]
\begin{tabular}{lc|c|c||c|c|c|c}
\hline
& \multicolumn{3}{c||}{state mixing times (ps)} & \multicolumn{4}{c}{structural correlation times (ps)} \\
& $\tau_2$ & $\tau_\mathrm{ac}$ & $\tau_\mathrm{end}$ & $\tau_{\cos \phi}$ & $\tau_{\sin \phi}$ & $\tau_{\cos \psi}$ & $\tau_{\sin \psi}$\\
\hline
neighbor exchange & 82 $\pm$ 4 & 31.0 $\pm$ 0.9 & 350 $\pm$ 30 & 47 $\pm$ 2 & 57 $\pm$ 2 & 26.4 $\pm$ 0.8 & 27.1 $\pm$ 0.9 \\
independence sampling \hspace{0.1in} & 24.2 $\pm$ 0.3 & 5.45 $\pm$ 0.06 & 175 $\pm$ 6 & 8.92 $\pm$ 0.09 & 9.9 $\pm$ 0.1 & 5.63 $\pm$ 0.04 & 6.09 $\pm$ 0.04 \\ % 2000 iterations
\hline
\end{tabular}
\caption{{\bf Efficiency measures for two-dimensional replica exchange umbrella sampling for the alanine dipeptide in implicit solvent.}
mixing times ins state space listed are: 
$\tau_2$, estimated from second eigenvalue of empirical state transition matrix; 
$\tau_\mathrm{ac}$, estimated from autocorrelation time of alchemical state index; 
$\tau_\mathrm{end}$, estimated average end-to-end transit time for alchemical state index.
Autocorrelation times of trigonometric functions of $\phi$ and $\psi$ torsion angles are listed as $\tau_{\cos \phi}, \tau_{\sin \phi}, \tau_{\cos \psi}, \tau_{\sin \psi}$.
The statistical error is given as one standard error of the mean.
\label{table:alanine-dipeptide-2d-umbrella-sampling}
}
\end{table*}

%%%%%%%%%%%%%%%%%%%%%%%%%%%%%%%%%%%%%%%%%%%%%%%%%%%%%%%%%%%%%
% DISCUSSION
%%%%%%%%%%%%%%%%%%%%%%%%%%%%%%%%%%%%%%%%%%%%%%%%%%%%%%%%%%%%%

\section{Discussion}
\label{section:discussion}

We have presented the framework of Gibbs sampling on the joint set of state and coordinate variables to better understand different expanded ensemble and replica exchange schemes, and demonstrated how this framework can identify simple ways to enhance the efficiency of expanded ensemble and replica exchange simulations by modifying the thermodynamic state update phase of the algorithms.
While the actual efficiency improvement will depend on the system and simulation details, we believe there is likely little, if any, drawback to using these improvements in a broad range of situations.

For simulated and parallel tempering simulations, in which only the temperature is varied among the thermodynamic states, the recommended scheme (independence sampling updates, Sections~\ref{section:algorithms:expanded-ensemble:gibbs-sampling} and \ref{section:algorithms:replica-exchange:gibbs-sampling}) is simple and inexpensive enough to be easily adopted by simulated and parallel tempering codes.
Because calculation of exchange probability requires no additional energy evaluations, it is effectively free.  
Other expanded ensemble or replica exchange simulations where the potential does not vary between states (such as exchange among temperatures and pressures~\cite{paschek-garcia:prl:2004:temperature-pressure-replica-exchange} or pH values~\cite{meng-roitberg:jctc:2010:constant-pH-replica-exchange}) are also effectively free, as no additional energy evaluations are required in these cases either. 
As long as state space evaluations are cheap compared to configuration updates, independence sampling will mix more rapidly than neighbor updates, though this advantage will be reduced as the interval spent between configuration updates by molecular dynamics or Monte Carlo simulation or the total time performing these coordinate updates becomes very small.

In some cases, exchange of information between processors during replica exchange in tightly coupled parallel codes may incur some cost, mainly in the form of latency.
In many cases, however, the decrease in mixing times could more than offset any loss in parallel efficiency.  
If the recommended independence sampling schemes would consume a substantial fraction of the iteration time, or where the parallel implementation of state updates is already complex, it may still be relatively inexpensive to simply perform the same state update scheme \emph{several times}, achieving enhanced mixing with little extra coding or computational overhead. Alternatively, the Gibbs sampling formalism could be used to design some other scheme that performs frequent state space sampling only on replicas that are local in the topology of the code. 

For simulated scaling~\cite{li-fajer-yang:jcp:2007:simulated-scaling} or Hamiltonian exchange simulations~\cite{sugita-kitao-okamoto:jcp:2000:hamiltonian-exchange,fukunishi-watanabe-takada:jcp:2002:hamiltonian-exchange,jang-shin-pak:prl:2003:hamiltonian-exchange,kwak-hansmann:prl:2005:hamiltonian-exchange}, independence sampling updates of state permutation vector $S$ requires evaluation of the reduced potential $u_k(x)$ at all $K$ states for the current configuration (in simulated scaling) or all replica configurations $x_k$ (for Hamiltonian exchange), which requires more energy evaluations than the neighbor exchange scheme.
However, if the intent is to make use of the multistate Bennett acceptance ratio (MBAR) estimator~\cite{shirts-chodera:jcp:2008:mbar}, which produces optimal estimates of free energy differences and expectations, all of these energies are required for analysis anyway, and so the computational impact on simulation time is negligible.
It is more computationally efficient to evaluate these additional reduced potentials \emph{during} the simulation, instead of post-processing simulation data, which is especially true if the additional reduced potential evaluations are done in parallel. Alternatively, if a simulated scaling simulation is run and one does not wish to use MBAR, restricted range state updates (Section~\ref{section:algorithms:expanded-ensemble:restricted-range-gibbs}) offer improved mixing behavior with minimal additional number of energy evaluations.

We have found that examining the exchange statistics, the empirical state transition matrix and its dominant eigenvalues, is extremely useful in diagnosing equilibration and convergence, as well as poor choices of thermodynamic states.
It is often very easy to see, from the diagonally dominant structure of this matrix, where regions of poor state overlap occur.
Poor overlap among sets of thermodynamic states observed early in simulations from the empirical state transition matrix are likely to also frustrate post-simulation analysis with techniques like MBAR and histogram reweighting methods~\cite{shirts-chodera:jcp:2008:mbar,chodera:jctc:2007:parallel-tempering-wham,kumars:WHAM}, making such metrics useful diagnostic tools.

For more complex state topologies in expanded ensemble or replica exchange simulations, where for example several different pressures or temperatures are included simultaneously, there may not exist a simple grid of values, or it may not be easy to identify which states are the most efficient neighbors. 
Using independence sampling eliminates the need to plan efficient exchange schemes among neighbors, or even to determine which states are neighbors.
This may encourage the addition of states that aid in reducing the correlation time of the overall Markov chain solely by speeding decorrelation of conformational degrees of freedom, since they will automatically couple to states with reasonable phase space overlap.

It is important to stress, however, that expanded ensemble and replica exchange simulations are not a cure-all for all systems with poor sampling.
In the presence of a first-order or pseudo-first-order phase transition, phase space mixing may still take an exponentially long time even when simulated or parallel tempering algorithms are used~\cite{bhatnagar-randall:acm:2004:torpid-mixing}.  
Optimization of the state exchange scheme, as described here, can only help so much; further efficiency gains would require design of intermediate states that abolish the first-order phase transition.
Schemes for optimal state selection are an area of active research~\cite{kofke:2002:jcp:acceptance-probability,katzberger-trebst-huse-troyer:j-stat-mech:2006:feedback-optimized-parallel-tempering,trebst-troyer-hansmann:jcp:2006:optimized-replica-selection,nadler-hansmann:pre:2007:generalized-ensemble,gront-kolinski:j-phys-condens-matter:2007:optimized-replica-selection,park-pande:pre:2007:choosing-weights-simulated-tempering,shenfeld-xu:pre:2009:thermodynamic-length}.

Finally, we observe that the independence sampling scheme for a simulated tempering simulation or any simulation where the contribution to the reduced potential is a thermodynamic parameter $\lambda$ multiplying a conjugate configuration-dependent variable $h(x)$ naturally generalizes to a continuous limit. As the number $K$ of thermodynamic states $\lambda_k$ is increased between some fixed lower and upper limits, this process eventually results in the thermodynamic state index $k$ effectively becoming a continuous variable $\lambda$~\cite{iba:intl-j-mod-phys-c:2001:extended-ensemble}.
Such a \emph{continuous tempering} simulation would sample from the joint distribution $\pi(x,\lambda) \propto \exp[-\lambda h(x) + g(\lambda)]$, with the continuous log weighting function $g(\lambda)$ replacing the discrete $g_k$ in simulated tempering simulations.

The Gibbs sampler and variations on it remain exciting areas for future exploration, and we hope that our conditional state space sampling formulation will make it much easier for other researchers to envision, develop, and implement new schemes for sampling from multiple thermodynamics states.  We also hope it encourages exploration of further connections between the two deeply interrelated fields of statistical mechanics and statistical inference.

%%%%%%%%%%%%%%%%%%%%%%%%%%%%%%%%%%%%%%%%%%%%%%%%%%%%%%%%%%%%%
% ACKNOWLEDGMENTS
%%%%%%%%%%%%%%%%%%%%%%%%%%%%%%%%%%%%%%%%%%%%%%%%%%%%%%%%%%%%%

\begin{acknowledgments}
The authors thank Sergio Bacallado (Stanford University), Jed Pitera (IBM Almaden), Adrian Roitberg (University of Florida), Scott Schmidler (Duke University), and William Swope (IBM Almaden) for insightful discussions on this topic, and Imran Haque (Stanford University), David Minh (Argonne National Labs), Victor Martin-Mayor (Universidad Complutense de Madrid), and Anna Schnider (UC-Berkeley) for a critical reading of the manuscript, and especially David M. Rogers (Sandia National Labs) who recognized a key error in an early version of the restricted range sampling method. JDC acknowledges support from a QB3-Berkeley Distinguished Postdoctoral Fellowship.  
Additionally, the authors are grateful to {\sc OpenMM} developers Peter Eastman, Mark Friedrichs, Randy Radmer, and Christopher Bruns and project leader Vijay Pande (Stanford University and SimBios) for their generous help with the {\sc OpenMM} GPU-accelerated computing platform and associated {\sc PyOpenMM} Python wrappers.
\end{acknowledgments}

%%%%%%%%%%%%%%%%%%%%%%%%%%%%%%%%%%%%%%%%%%%%%%%%%%%%%%%%%%%%%
% BIBLIOGRAPHY
%%%%%%%%%%%%%%%%%%%%%%%%%%%%%%%%%%%%%%%%%%%%%%%%%%%%%%%%%%%%%

\bibliography{gibbs-sampling}

\end{document}